\documentclass[12pt]{article}
\usepackage{a4wide,amssymb}
\pdfoutput=1
\usepackage{a4wide,amssymb,graphicx}
\usepackage{epsfig}
\usepackage[usenames,dvipsnames]{color}
\usepackage{slashed}
\newcommand{\be}{\begin{equation}}
\newcommand{\ee}{\end{equation}}
\newcommand{\bea}{\begin{eqnarray}}
\newcommand{\eea}{\end{eqnarray}}

\def\circa#1{\,\raise.3ex\hbox{$#1$\kern-.75em\lower1ex\hbox{$\sim$}}\,}

\begin{document}

\begin{titlepage}
%
%


%

\begin{centering}
\vspace{1cm}
{\Large {\bf Diphoton resonance confronts dark matter}} \\

\vspace{1.5cm}

{\bf Soo-Min Choi$^*$, Yoo-Jin Kang$^\dagger$   and  Hyun Min Lee$^\ddagger$}
\vspace{.5cm}

{\it Department of Physics, Chung-Ang University, 06974 Seoul, Korea.} 
\\

\end{centering}
\vspace{2cm}

\begin{abstract}
\noindent
As an interpretation of the 750 GeV diphoton excesses recently reported by both ATLAS and CMS collaborations, we consider a simple extension of the Standard Model with a Dirac fermion dark matter where a singlet complex scalar field mediates between dark matter and SM particles via
effective couplings to SM gauge bosons and/or Higgs-portal. In this model, we can accommodate the diphoton events through the direct and/or cascade decays of pseudo-scalar and real scalar partners of the complex scalar field. We show that mono-jet searches and gamma-ray observations are complementary in constraining the region where the width of the diphoton resonance can be enhanced due to the couplings of the resonance to dark matter and the correct relic density is obtained. 
In the case of cascade decay of the resonance, the effective couplings of singlet scalars can be smaller, but the model is still testable by the future discrimination between single photon and photon-jet at the LHC as well as the gamma-ray searches for the cascade annihilation of dark matter.

\end{abstract}

\vspace{3cm}

\begin{flushleft}
$^*$Email: sm90515@cau.ac.kr \\
$^\dagger$Email: dbwlsdl1008@cau.ac.kr \\
$^\ddagger$Email: hminlee@cau.ac.kr 
\end{flushleft}

\end{titlepage}

\section{Introduction}

Recently there have been tantalizing hints for new physics from the diphoton excesses at about $750\,{\rm GeV}$ with local significances of $3.9\sigma$ and $2.6\sigma$, that have been observed in LHC Run 1 data at $13\,{\rm TeV}$ by both ATLAS and CMS collaborations \cite{diphoton}, respectively. After Moriond 2016 conference, the results are updated with $8\,{\rm TeV}$ data included in the analysis \cite{moriond}, leading to higher significances, in particular, to $3.4\sigma$ in the case of CMS. 
ATLAS favors a wide width of the resonance about $45\,{\rm GeV}$, but the significance changes only by $0.3\sigma$ as compared to the case with narrow width. Furthermore, CMS prefers a narrow width in their best fit result. The production cross section required for explaining the diphoton excesses is about $6\,{\rm fb}$, although the result depends on the assumption of the resonance width \cite{scalar,production}. 
The production cross section for the diphoton resonance appears relatively large for given collider bounds from other related LHC searches \cite{scalar}. The typical interpretation of diphoton excesses with a new scalar resonance calls for extra vector-like fermions with sizable Yukawa couplings to the resonance \cite{scalar,scalar2}. In the case of a spin-2 resonance such as Kaluza-Klein graviton, a nontrivial positioning of SM particles in extra dimensions is necessary to satisfy strong bounds from electroweak precision data and dilepton and di-jet searches \cite{gmdm-diphoton,spin2}. 
Unitarity arguments on the resonance might imply the coexistence of scalar and extra resonances of higher spin in QCD-like theories or gravity duals in extra dimensions \cite{coexist}.


At the moment, we don't have enough information to tell about the properties of  the resonance such as width and spin/parity, but we will be able to know them from LHC Run 2 data at $13\,{\rm TeV}$.
In the mean time, it would be interesting to entertain the possibility of a sizable width scenario that can be consistent with the diphoton excesses and other experimental bounds. If the invisible decay mode of the resonance, which is less constrained, is responsible for a large width of the resonance, there is an interesting possibility that the resonance plays a role of mediator between the SM and dark matter \cite{gmdm-diphoton,diphotondm}. On the other hand, there is a plausible option to explain the diphoton excesses with collimated photons, the so called photon-jets, which come from a cascade decay of the resonance into a pair of light mediators, each of them decaying into a pair of photons \cite{cascade}.  In this case, the width of the resonance can be increased by a renormalizable coupling between the resonance and the light mediator.


In this article, we consider a simple extension of the SM with a complex singlet scalar field that couples to both the SM and Dirac fermion dark matter in the presence of an approximate $U(1)$ global symmetry. A soft breaking of the global symmetry induces a nonzero mass for the would-be Goldstone boson or pseudo-scalar, so the model is consistent with phenomenological bounds. We introduce effective couplings of real-scalar and pseudo-scalar of the complex scalar field to the SM gauge bosons as a consequence of integrating out new vector-like fermions and the real-scalar can also couple to the SM particles just like the SM Higgs via Higgs-portal. The $U(1)$ invariant couplings of the complex scalar field to vector-like fermions fix the ratio of effective couplings of real-scalar and pseudo-scalar in our model.  

We identify the real-scalar and/or pseudo-scalar as the diphoton resonance in our model and consider the possibilities of explaining the diphoton excesses in terms of the direct and/or cascade decays of the resonance. In each case, we impose the collider bounds such as mono-jet and di-jet bounds as well as indirect bounds from gamma-ray and anti-proton searches for dark matter. 
As illustrated from benchmark models that satisfy all the phenomenological constraints, we show that there is an interesting interplay between mono-jet and gamma-ray searches in the case of direct decay whereas those bounds can be weakened in the case of cascade decay due to smaller effective couplings of the singlet scalars. In the latter case, the discrimination between single photon and photon-jet in the LHC Run 2 would become more important. On the side of cosmic data, the same coupling responsible for the cascade decay of the resonance leads to the cascade annihilation of dark matter into multiple photons leading to interesting signatures such as gamma-ray box.

This paper is organized as follows. 
We begin with a description for the interactions of singlet scalars in our model and present the necessary formulas for the partial decay rates. Then, we discuss the diphoton conditions in the cases of direct and/or cascade decays of the singlet scalar(s) and constrain the parameter space of effective couplings of the resonances. 
Next we consider the annihilation of dark matter with the singlet scalar mediators in each scenario of the diphoton interpretation and show how collider and cosmic data can be used to constrain the models.  There is an appendix containing the scalar potential and scalar self-interactions in our model. 
Finally, conclusions are drawn.

\section{The model}

We consider a complex singlet scalar $S$ and a Dirac fermion dark matter $\chi$, that transform under a $U(1)$ global symmetry as $S\rightarrow e^{-2i\alpha} S$ and $\chi\rightarrow e^{i\gamma_5\alpha}\chi$, respectively. 
Expanding the complex scalar $S$ around a VEV as in the appendix  and integrating out vector-like fermions \cite{lpp,MET,ibarra}, we obtain the following effective Lagrangian for a singlet pseudo-scalar  $a$, two CP-even scalars, Higgs-like $h_1$ and singlet-like $h_2$, and dark matter,
\bea
{\cal L}&=&{\bar \chi} (i\gamma^\mu \partial_\mu-m_\chi) \chi+\frac{1}{2}(\partial_\mu a)^2-\frac{1}{2}m^2_a a^2+\sum_{i=1,2}\bigg(\frac{1}{2}(\partial_\mu h_i)^2-\frac{1}{2}m^2_i h^2_i\bigg) \\
&&+\frac{1}{\sqrt{2}}\,i \lambda_\chi a \,{\bar \chi}\gamma^5 \chi -\frac{1}{\sqrt{2}} \lambda_\chi (h_2\cos\theta+h_1\sin\theta) {\bar\chi}\chi +{\cal L}_{\rm scalar}+{\cal L}_{\rm D5} \label{effaction}
\eea
where ${\cal L}_{\rm scalar}$ is the interaction Lagrangian for scalars only given in eq.~(\ref{scalar}) and ${\cal L}_{D5}$ contains the dimension-5 interactions of singlet scalars to gauge fields, given by
\bea
{\cal L}_{\rm D5}&=&\frac{1}{\Lambda}\, a \Big(c_1 F^Y_{\mu\nu}{\tilde F}^{ Y\mu\nu}+c_2 W_{\mu\nu} {\tilde W}^{\mu\nu}+c_3 G_{\mu\nu} {\tilde G}^{\mu\nu}\Big)  \nonumber \\
&&+\frac{1}{\Lambda}\, h_2\cos\theta \Big(d_1 F^Y_{\mu\nu}{F}^{ Y\mu\nu}+d_2 W_{\mu\nu} {W}^{\mu\nu}+d_3 G_{\mu\nu} {G}^{\mu\nu}\Big)  \nonumber \\
&&+\frac{1}{\Lambda}\, h_1\sin\theta \Big({\hat d}_1 F^Y_{\mu\nu}{F}^{ Y\mu\nu}+{\hat d}_2 W_{\mu\nu} {W}^{\mu\nu}+{\hat d}_3 G_{\mu\nu} {G}^{\mu\nu}\Big)  \label{loops}
\eea
with the dual field strength tensor being ${\tilde F}_{\mu\nu}\equiv \frac{1}{2}\epsilon_{\mu\nu\rho\sigma} F^{\rho\sigma}$, etc, and  $c_i, d_i, {\hat d}_i (i=1,2,3)$ being effective couplings induced by vector-like fermions. We note that when vector-like fermion have the same global charges as for dark matter \cite{lpp}, in the decoupling limit of vector-like fermions, the effective couplings are related by $d_i=\frac{4}{3} c_i(i=1,2,3)$.  
Then, singlet scalars communicate between dark matter and the SM particles, via effective gauge couplings and Higgs-portal. Similar models \cite{lpp,hambye,MET} have been considered in light of the Fermi-LAT gamma-ray line, satisfying various bounds from indirect and direct detections as well as collider experiments. 
The interplay between dark matter detection and collider experiments in the cases with vector \cite{MET} or tensor \cite{gmdm} mediators have been also discussed in the previous works.

In our model, the $U(1)$ global symmetry is broken to a $Z_2$ discrete symmetry, which ensures the stability of dark matter fermion. Any global charges are vulnerable to quantum gravity effects \footnote{See a recent discussion on the classification of effective interactions that violate the global symmetries of dark matter \cite{global}. }, but the violation of a global symmetry could be ensured at sufficiently higher orders of effective interactions in the presence of extra discrete gauge symmetries \cite{discrete}. We assume that this is the case without changing the low-energy phenomenology.

In the basis of physical gauge bosons, the loop-induced couplings in eq.~(\ref{effaction}) can be rewritten as
\bea
{\cal L}_{\rm D5}&=&\frac{1}{\Lambda}\, a \Big(c_{\gamma\gamma} F_{\mu\nu}{\tilde F}^{ \mu\nu}+c_{\gamma Z} F_{\mu\nu} {\tilde Z}^{\mu\nu}+c_{WW} W^+_{\mu\nu} {\tilde W}^{-\mu\nu}+c_{ZZ} Z_{\mu\nu} {\tilde Z}^{\mu\nu}+c_{gg} G_{\mu\nu} {\tilde G}^{\mu\nu}\Big)  \nonumber \\
&&+\frac{1}{\Lambda}\, h_2\Big(d_{\gamma\gamma} F_{\mu\nu}{ F}^{ \mu\nu}+d_{\gamma Z} F_{\mu\nu} { Z}^{\mu\nu}+d_{WW} W^+_{\mu\nu} { W}^{-\mu\nu} \nonumber \\
&&\quad\quad +d_{ZZ} Z_{\mu\nu} {Z}^{\mu\nu}+d_{gg} G_{\mu\nu} { G}^{\mu\nu}\Big)  \nonumber \\
&&+\frac{1}{\Lambda}\, h_1\Big({\hat d}_{\gamma\gamma} F_{\mu\nu}{ F}^{ \mu\nu}+{\hat d}_{\gamma Z} F_{\mu\nu} { Z}^{\mu\nu}+{\hat d}_{WW} W^+_{\mu\nu} { W}^{-\mu\nu} \nonumber \\
&&\quad\quad +{\hat d}_{ZZ} Z_{\mu\nu} {Z}^{\mu\nu}+{\hat d}_{gg} G_{\mu\nu} { G}^{\mu\nu}\Big)
\label{loops2}
\eea
with
\bea
c_{\gamma\gamma}&=&c_1\cos^2\theta_W +c_2 \sin^2\theta_W,  \quad\quad  c_{\gamma Z}= (c_2-c_1)\sin(2\theta_W), \\
c_{WW}&=& 2c_2. \quad\quad  c_{ZZ}= c_1\sin^2\theta_W +c_2 \cos^2\theta_W, \quad\quad  c_{gg}=c_3,   \\
d_{\gamma\gamma}&=&(d_1\cos^2\theta_W +d_2 \sin^2\theta_W)\cos\theta,  \quad\quad  d_{\gamma Z}= (d_2-d_1)\sin(2\theta_W)\cos\theta, \\
d_{WW}&=& 2d_2\cos\theta. \quad\quad  d_{ZZ}= (d_1\sin^2\theta_W +d_2 \cos^2\theta_W)\cos\theta, \quad\quad  d_{gg}=d_3\cos\theta,
\eea
and
\bea
{\hat d}_{\gamma\gamma}&=&({\hat d}_1\cos^2\theta_W +{\hat d}_2 \sin^2\theta_W)\sin\theta,  \quad\quad  {\hat d}_{\gamma Z}= ({\hat d}_2-{\hat d}_1)\sin(2\theta_W)\sin\theta, \\
{\hat d}_{WW}&=& 2{\hat d}_2\sin\theta. \quad\quad {\hat d}_{ZZ}= ({\hat d}_1\sin^2\theta_W +{\hat d}_2 \cos^2\theta_W)\sin\theta, \quad\quad  {\hat d}_{gg}={\hat d}_3\sin\theta.
\eea

The total decay rate of the pseudo-scalar is given by $\Gamma_a=\sum_i \Gamma_a(i)$, with the partial decay rates of the pseudo-scalar being
\bea
\Gamma_a(gg) &=&  \frac{2m^3_a}{\pi\Lambda^2 } \,c^2_{gg}, \label{agg} \\
\Gamma_a(\gamma\gamma)&=& \frac{m^3_a}{4\pi\Lambda^2 } \,c^2_{\gamma\gamma}, \label{arr} \\
\Gamma_a(Z\gamma) &=& \frac{m^3_a}{8\pi \Lambda^2}\,c^2_{Z\gamma} \Big(1-\frac{m^2_Z}{m^2_a}\Big)^3, \\
\Gamma_a(ZZ) &=&  \frac{m^3_a}{4\pi\Lambda^2 }\,c^2_{ZZ} \Big(1-\frac{4m^2_Z}{m^2_a}\Big)^{3/2}, \\
\Gamma_a(WW)&=&  \frac{m^3_a}{8\pi \Lambda^2} \,c^2_{WW} \Big(1-\frac{4m^2_W}{m^2_a}\Big)^{3/2} \\
\Gamma_a({\bar\chi}\chi) &=& \frac{\lambda^2_\chi m_a}{16\pi}\,\Big(1-\frac{4m^2_\chi}{m^2_a}\Big)^{1/2}.
\eea

The case with a nonzero Higgs mixing angle is potentially interesting for the exotic decays of the SM Higgs boson. However, given the strong limits from Higgs data \cite{atlasH,cmsH,atlasg2,cmsg2}, we focus on the case with a negligibly small Higgs mixing, $\sin\theta\sim 0$, so that there is no modification in the couplings of the SM Higgs. For $\sin\theta\sim 0$, the total decay rate of the singlet-like scalar is given by $\Gamma_2=\sum_i \Gamma_{h_2}(i)$, with the partial decay rates of the pseudo-scalar being
\bea
\Gamma_2(gg) &=&  \frac{2m^3_2}{\pi\Lambda^2 } \,d^2_{gg},  \label{sgg}\\
\Gamma_{2}(\gamma\gamma)&=& \frac{m^3_2}{4\pi\Lambda^2 } \,d^2_{\gamma\gamma},  \label{srr}\\
\Gamma_{2}(Z\gamma) &=& \frac{m^3_2}{8\pi \Lambda^2}\,d^2_{Z\gamma} \Big(1-\frac{m^2_Z}{m^2_2}\Big)^3, \\
\Gamma_{2}(ZZ) &=&  \frac{m^3_2}{4\pi\Lambda^2 }\,d^2_{ZZ}\Big(1-\frac{4m^2_Z}{m^2_2}+\frac{6m^4_Z}{m^4_2}\Big) \Big(1-\frac{4m^2_Z}{m^2_2}\Big)^{1/2}, \\
\Gamma_{2}(WW)&=&  \frac{m^3_2}{8\pi \Lambda^2} \,d^2_{WW}\Big(1-\frac{4m^2_W}{m^2_2}+\frac{6m^4_W}{m^4_2}\Big) \Big(1-\frac{4m^2_W}{m^2_2}\Big)^{1/2}, \\
\Gamma_{2}({\bar\chi}\chi) &=& \frac{\lambda^2_\chi m_2}{16\pi}\Big(1-\frac{4m^2_\chi}{m^2_2}\Big)^{3/2}, \\
\Gamma_{2}(aa)&=& \frac{\lambda^2_S v^2_s}{8\pi m_{2}} \Big(1-\frac{4m^2_a}{m^2_{2}} \Big)^{1/2}.
\eea
Here, in the $aa$ decay mode, $\lambda_S$ is the quartic coupling for the complex scalar field $S$, as introduced in the appendix.

\section{Diphoton resonance at the LHC}

The recently observed diphoton excess near $750\,{\rm GeV}$ can be explained by the direct decay of a new neutral resonance beyond the SM. Nonetheless, there are more possibilities to explain the diphoton resonance with the direct decays of two degenerate resonances or with a cascade decay of the resonance into multi-photons through light intermediate states. 
In this section, we consider each of the possibilities in the model and constrain the effective couplings of the resonances.

\begin{figure}
  \begin{center}
   \includegraphics[height=0.20\textwidth]{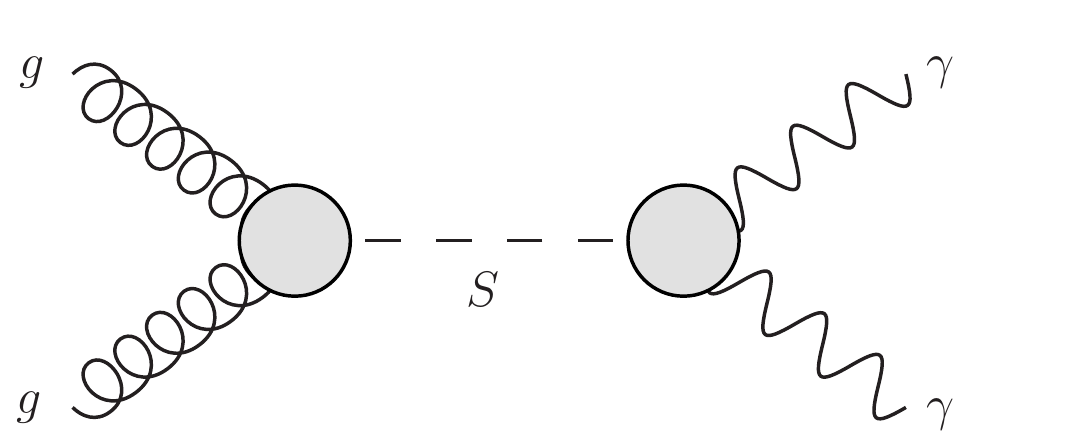}
    \includegraphics[height=0.20\textwidth]{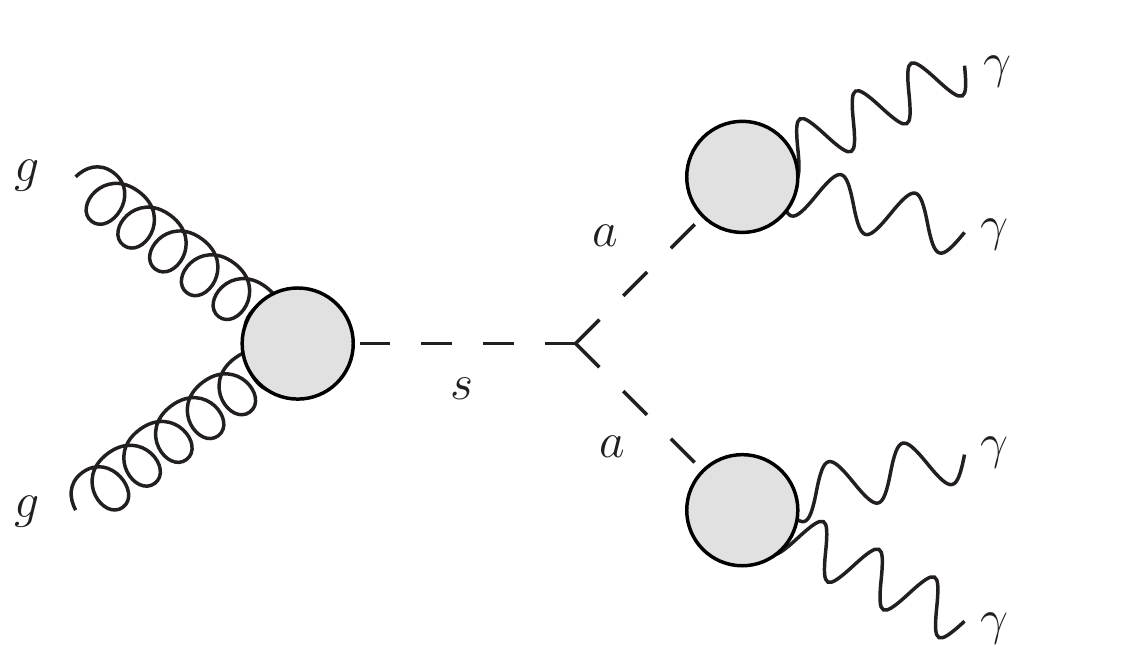}
   \end{center}
  \caption{Feynman diagrams for diphoton production with gluon fusion.}
  \label{diphoton-diagrams}
\end{figure}

\subsection{Diphotons from direct decay}

The resonance production cross section of scalar particle(s) $X$ via gluon fusion followed by its diphoton decay at the LHC is given \cite{scalar} by
\bea
\sigma(pp\rightarrow X\rightarrow\gamma\gamma)=\frac{1}{sM_X \Gamma_X}\,K_{gg} C_{gg} \Gamma(X\rightarrow gg) \Gamma(X\rightarrow\gamma\gamma) 
\eea
where the gluon luminosity is given by $C_{gg}=2137$ for $s=(13\,{\rm TeV})^2$ and the $K$-factor is given by $K_{gg}=1.5$.  In the case with resonances from pseudo-scalar $a$ and/or real scalar $s$, we have $X=a$ and/or $s$.

\begin{figure}
  \begin{center}
   \includegraphics[height=0.33\textwidth]{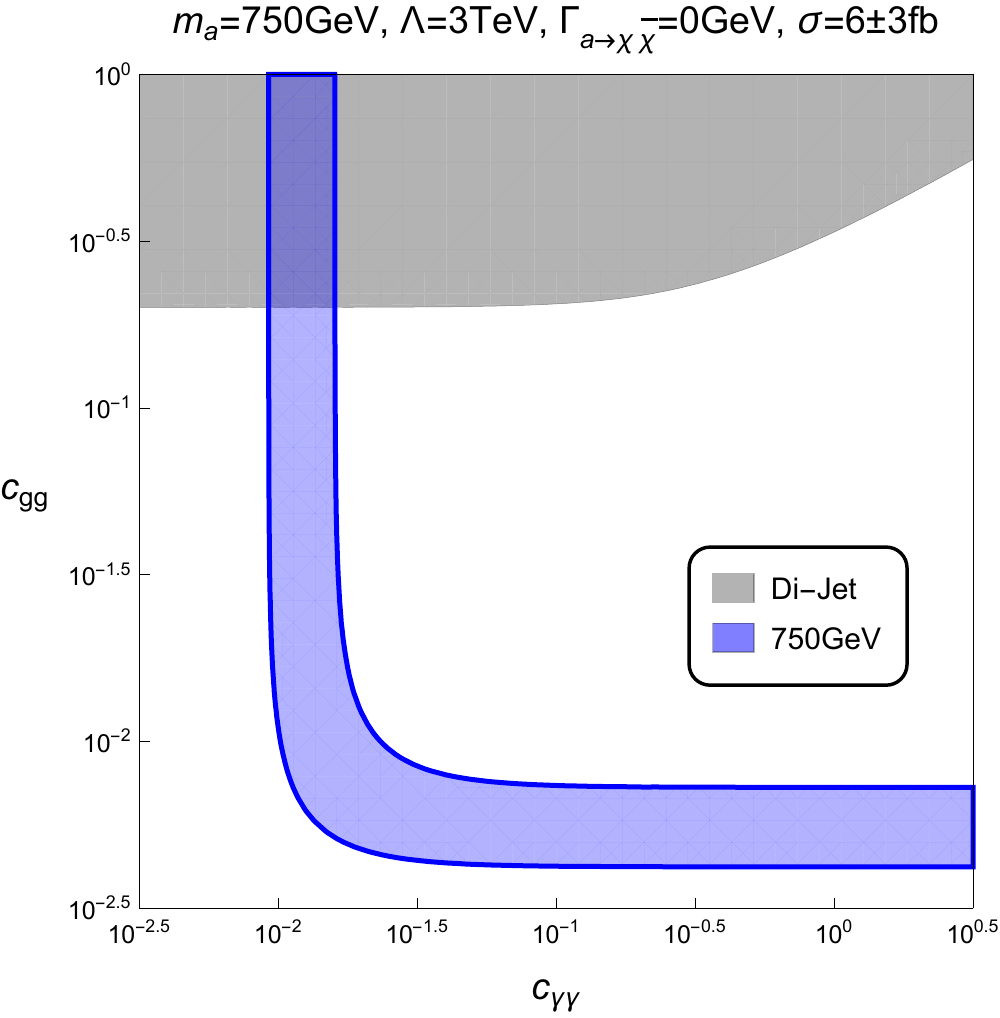}
    \includegraphics[height=0.33\textwidth]{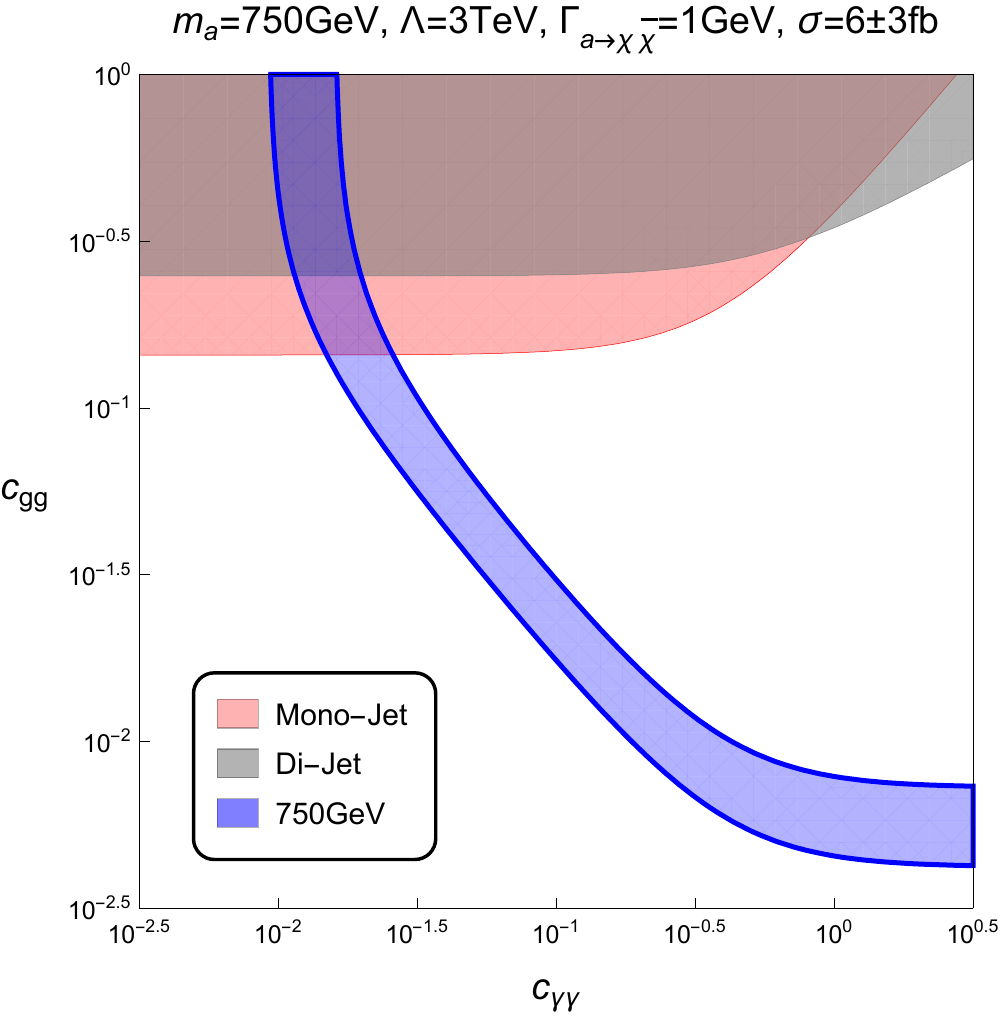} 
    \includegraphics[height=0.33\textwidth]{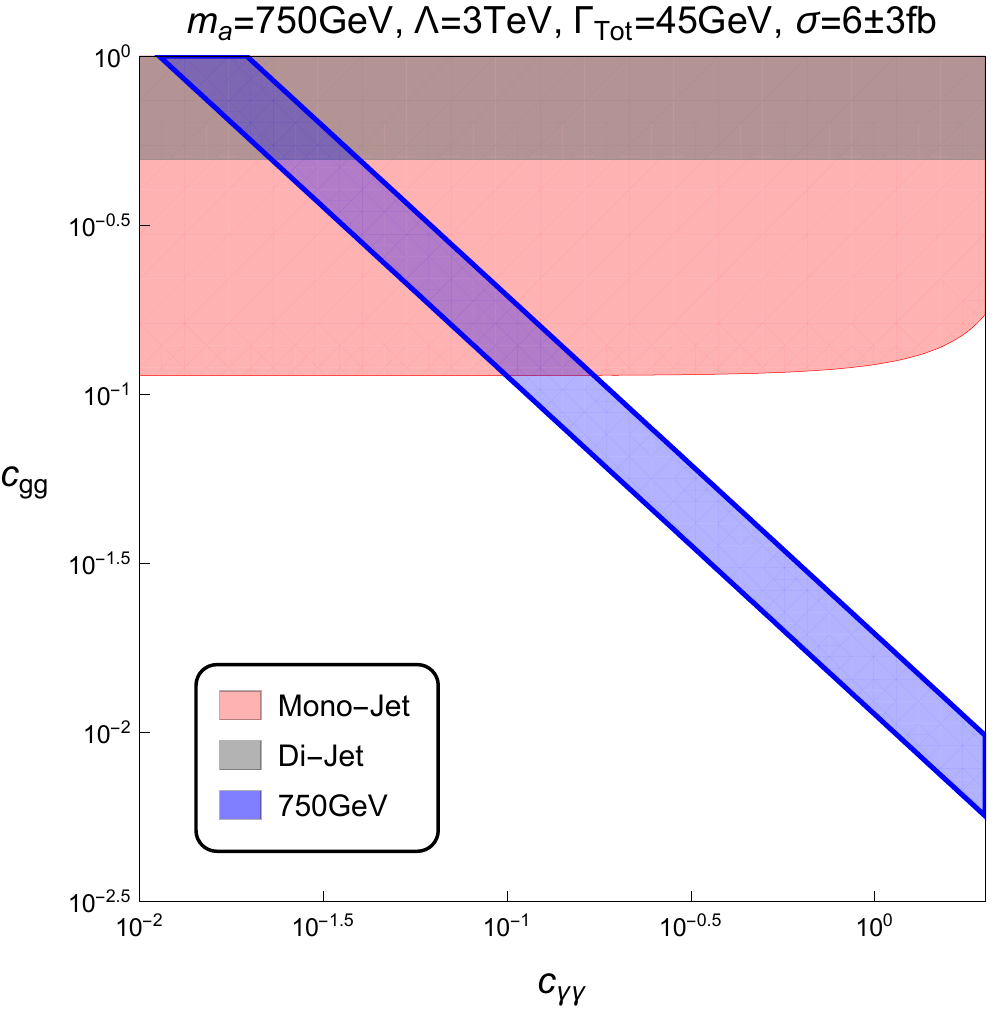}
   \end{center}
  \caption{Parameter space of $c_{\gamma\gamma}$ and $c_{gg}$ for the diphoton resonance, from the decay of the pseudo-scalar. We have taken $c_2=0$, $c_1\neq 0$ and $c_3\neq 0$ in the unbroken phase. The region explaining the diphoton resonance at $750\,{\rm GeV}$ for $\sigma(pp\rightarrow\gamma\gamma)=6\pm 3\,{\rm fb}$ is shown in blue strip.
   The regions excluded by mono-jet and di-jet limits from LHC $8\,{\rm TeV}$ are shown in pink (less dark) and gray (dark), respectively. The invisible decay width of the pseudo-scalar is chosen to $\Gamma_{\rm inv}=0, 1\,{\rm GeV}$ in the left-most and middle plots, respectively, and the total decay rate is $\Gamma_{\rm tot}=45\,{\rm GeV}$ in the right-most plot. The case with real-scalar resonance is similar. }
  \label{couplings}
\end{figure}

When there is only one resonance due to either pseudo-scalar or real scalar, i.e., $X=a$ or $s$, the diphoton production cross section leads to
\bea
\Gamma(X\rightarrow gg) \Gamma(X\rightarrow \gamma\gamma)=2.8\times 10^{-2}{\rm GeV}^2  \Big(\frac{\sigma(pp\rightarrow \gamma\gamma)}{6\,{\rm fb}} \Big)\Big(\frac{\Gamma_X}{45\,{\rm GeV}}\Big). 
\eea
For $X=a$, from eqs.~(\ref{agg}) and (\ref{arr}), we get the condition on the effective couplings of the resonance as
\bea
|c_{gg}\cdot c_{\gamma\gamma}| = 0.016 \Big(\frac{\sigma(pp\rightarrow \gamma\gamma)}{6\,{\rm fb}} \Big)^{1/2} \Big(\frac{\Lambda}{3\,{\rm TeV}}\Big)^2 \Big(\frac{\Gamma_a}{45\,{\rm GeV}} \Big)^{1/2}.
\label{diphoton-a}
\eea
The $gg$ and $\gamma\gamma$ modes only do not tend to give rise to a wide width due to di-jet bound, unless the photon coupling to the resonance is large.  Therefore, if the wide width of the resonance is necessary, one has to rely on other decay modes of the resonance, such as the invisible decay mode into a pair of dark matter particles. In Fig.~\ref{couplings}, we depict the parameter space for the effective gluon and photon couplings of the pseudo-scalar field, explaining the diphoton excess and satisfying the di-jet bound as well as the mono-jet bound in the presence of the invisible decay mode. We have set $c_2=0$ and the diphoton production cross section of $\sigma(pp\rightarrow\gamma\gamma)=6\pm 3\,{\rm fb}$ is imposed.
 As we increase the invisible decay rate, the mono-jet bound becomes more sensitive to rule out a sizable gluon coupling.  

For $X=h_2$, there is a similar condition for the diphoton resonance, with $c_{gg}, c_{\gamma\gamma}$ being replaced by $d_{gg}, d_{\gamma\gamma}$, respectively, so there are similar limits from mono-jet and di-jet searches as those obtained for the pseudo-scalar resonance in the later discussion.

When there are two resonances with $m_a\approx m_2\approx 750\,{\rm GeV}$, namely, $X=a$ and $h_2$, two singlet scalars contribute to the diphoton excesses, with
the diphoton production cross section being constrained by
\bea
\Gamma_a(gg)\Gamma_a( \gamma\gamma)+\frac{\Gamma_a}{\Gamma_2}\,\Gamma_2(gg)\Gamma_2( \gamma\gamma)=2.8\times 10^{-2}{\rm GeV}^2 \Big(\frac{\sigma(pp\rightarrow \gamma\gamma)}{6\,{\rm fb}} \Big) \Big(\frac{\Gamma_a}{45\,{\rm GeV}}\Big). 
\eea
Then, from eqs.~(\ref{agg}), (\ref{arr}), (\ref{sgg}) and (\ref{srr}), the effective couplings of the resonances are constrained to
\bea
\sqrt{c^2_{gg} c^2_{\gamma\gamma}+\frac{\Gamma_a}{\Gamma_2}\, d^2_{gg} d^2_{\gamma\gamma} }=0.016 \Big(\frac{\sigma(pp\rightarrow \gamma\gamma)}{6\,{\rm fb}} \Big)^{1/2} \Big(\frac{\Lambda}{3\,{\rm TeV}}\Big)^2 \Big(\frac{\Gamma_a}{45\,{\rm GeV}} \Big)^{1/2}.
\eea
In the decoupling limit of vector-like fermions that have the same global charges as dark matter, we get $d_{gg}=\frac{4}{3} c_{gg}$ and $d_{\gamma\gamma}=\frac{4}{3} c_{\gamma\gamma}$ again so the above condition becomes
\bea
|c_{gg}\cdot c_{\gamma\gamma}|  \sqrt{1+\frac{256\Gamma_a}{81\Gamma_2}}= 0.016 \Big(\frac{\sigma(pp\rightarrow \gamma\gamma)}{6\,{\rm fb}} \Big)^{1/2} \Big(\frac{\Lambda}{3\,{\rm TeV}}\Big)^2 \Big(\frac{\Gamma_a}{45\,{\rm GeV}} \Big)^{1/2}.  \label{diphoton-deg}
\eea
In this case, the required values for the effective couplings of the pseudo-scalar can be weaker than the case with pseudo-scalar resonance only.  When scalars decay only into a pair of SM gauge bosons, the left-hand side in eq.~(\ref{diphoton-deg}) becomes $\sqrt{2}|c_{gg}\cdot c_{\gamma\gamma}|$, so the required effective couplings are reduced accordingly.

\begin{table}[ht]
\centering
\small
\begin{tabular}{|c||c|c|c|c|c|c|}
\hline 
Model & ${\rm BR}_a(\gamma\gamma)$ & ${\rm BR}_a(gg)$ & ${\rm BR}_a(Z\gamma)$ & ${\rm BR}_a(ZZ)$ & ${\rm BR}_a(\chi\bar{\chi})$ & $\Gamma_{a,{\rm tot}}[{\rm GeV}]$   \\ [0.7ex] 
\hline
A & $5.31\times 10^{-4}$ & $0.840$ & $3.07\times 10^{-4}$ & $4.36\times 10^{-5}$ & $0.159$ & $1.42$ \\ [0.7ex]
\hline
B & $2.12\times 10^{-3}$ & $0.016$ & $1.23\times 10^{-3}$ & $1.74\times 10^{-4}$ & $0.980$ & $18.6$ \\ [0.7ex]
\hline
C & $2.12\times 10^{-3}$ & $0.996$ & $1.23\times 10^{-3}$ & $1.74\times 10^{-4}$ & $-$ & $0.30$ \\ [0.7ex]
\hline
\end{tabular} 
\caption{Decay branching fractions and total decay rate of pseudo-scalar resonance. 
Benchmark models with $(c_{gg}, m_\chi, \lambda_\chi)$ are Model A : $(0.2,347 \,{\rm GeV} ,0.2)$; 
Model B : $(0.1,293\,{\rm GeV} ,1.4)$;
Model C : $(0.1,800\,{\rm GeV} ,1.8)$. The diphoton condition (\ref{diphoton-a}) with $\sigma(pp\rightarrow\gamma\gamma)=6\,{\rm fb}$ leads to $c_{\gamma\gamma}=0.0142, 0.103, 0.0130$, in the order of models. We have taken $c_2=0$, $c_1\neq 0$ and $c_3\neq 0$ in the unbroken phase. For all models, we have taken $m_a=750\,{\rm GeV}$ and the current collider bounds  are fullfilled. These benchmark models will be used for dark matter discussion in Table~\ref{ann-pseudo} in Section 4.}
\label{BRa}
\end{table}  

We remark on the important collider bounds on the model from the LHC. 
First, the mono-jet bound from CMS $8\,{\rm TeV}$ \cite{monojet} is given by
\be
\sigma(pp\rightarrow X\rightarrow \chi {\bar\chi}) < 0.8\,{\rm pb}
\ee
which is translated  to the bound on the ratio of the partial decays at LHC $13\,{\rm TeV}$,
\be
\frac{\Gamma(a\rightarrow \chi{\bar\chi})}{\Gamma(a\rightarrow \gamma\gamma)}< 667\,\Big(\frac{r}{5}\Big) \Big(\frac{6\,{\rm fb}}{\sigma(pp\rightarrow\gamma\gamma )}\Big)
\ee
where $r$ is the parton luminosity ratio given by $r=(C_{gg}/s)_{\rm 13\,{\rm TeV}}/(C_{gg}/s)_{8\,{\rm TeV}}\simeq 4.7$. 
Then, for $r=4.7$ and $\sigma(pp\rightarrow\gamma\gamma)=6\,{\rm fb}$, the mono-jet bound constrains the dark matter coupling to the resonance as
\be
\frac{\lambda^2_\chi}{4c^2_{\gamma\gamma}}\frac{\Lambda^2}{m_a^2}\lesssim 627. 
\ee
For $\Lambda=3\,{\rm TeV}$ and $m_a=750\,{\rm GeV}$, we get 
\be
|\lambda_\chi| \lesssim  13|c_{\gamma\gamma}|. \label{monojet}
\ee
For $|c_{\gamma\gamma}|={\cal O}(1)$, the mono-jet bound does not constrain the dark matter coupling much, but the case is strongly limited by indirect detection such as Fermi-LAT gamma-ray searches as will be discussed in the next section.

\begin{table}[ht]
\centering
\small
\begin{tabular}{|c||c|c|c|c|c|c|}
\hline 
Model & ${\rm BR}_s(\gamma\gamma)$ & ${\rm BR}_s(gg)$ & ${\rm BR}_s(Z\gamma)$ & ${\rm BR}_s(ZZ)$ & ${\rm BR}_s(\chi\bar{\chi})$ & $\Gamma_{s,{\rm tot}}[{\rm GeV}]$   \\ [0.7ex] 
\hline 
A & $5.31\times 10^{-4}$ & $0.941$ & $3.07\times 10^{-4}$ & $4.36\times 10^{-5}$ & $5.83\times 10^{-2}$ & $1.27$ \\ [0.7ex]
\hline
B & $5.31\times 10^{-4}$ & $0.999$ & $3.07\times 10^{-4}$ & $4.36\times 10^{-5}$ & $-$ & $1.19$ \\ [0.7ex]
\hline
C & $0.0785$ & $1.70\times 10^{-4}$ & $0.0438$ & $6.03\times 10^{-3}$ & $0.872$ & $47.5$ \\ [0.7ex]
\hline
\end{tabular}
\caption{Decay branching fractions and total decay rate of real-scalar resonance.  
Benchmark models are Model A : $(d_{gg}, m_\chi, \lambda_\chi)=(0.2, 361\,{\rm GeV} ,0.5)$; 
Model B : $(d_{gg}, m_\chi, \lambda_\chi)=(0.2,800\,{\rm GeV} ,1.4)$;
Model C : $(d_{\gamma\gamma}, m_\chi, \lambda_\chi)=(1.0,265\,{\rm GeV} ,2.8)$. 
The diphoton condition (\ref{diphoton-a}) for $\sigma(pp\rightarrow\gamma\gamma)=6\,{\rm fb}$ with $c_{gg}, c_{\gamma\gamma}$ being replaced by $d_{gg}, d_{\gamma\gamma}$, respectively,   leads to $d_{\gamma\gamma}=0.0134, 0.0130$, $d_{gg}=0.0165$, in the order of models. 
We have taken $c_2=0$, $c_1\neq 0$ and $c_3\neq 0$ in the unbroken phase. For all models, we have taken $m_s=750\,{\rm GeV}$ and the current collider bounds are fulfilled. These benchmark models will be used for dark matter discussion in Table~\ref{ann-real} in Section 5.}
\label{BRs}
\end{table}  

Furthermore, the di-jet bound at  LHC 8 TeV, $\sigma(pp\rightarrow X\rightarrow jj)<2.5{\rm pb}$ \cite{di-jet}, constrains the pseudo-scalar couplings by
\bea
\frac{\Gamma(a\rightarrow gg)}{\Gamma(a\rightarrow \gamma\gamma)} \lesssim 2083  \,\Big(\frac{r}{5}\Big) \Big(\frac{6\,{\rm fb}}{\sigma(pp\rightarrow\gamma\gamma )}\Big).
\eea
Thus, for $r=4.7$ and $\sigma(pp\rightarrow\gamma\gamma)=6\,{\rm fb}$, the di-jet bound constrains the  gluon coupling to the resonance as
\bea
|c_{gg}| \lesssim 15.6|c_{\gamma\gamma}|.  \label{di-jet}
\eea
When the real scalar is the diphoton resonance, a similar bound on the gluon coupling $d_{gg}$ applies. 

In Tables~\ref{BRa} and \ref{BRs}, we show the branching fractions and total decay rates of the pseudo-scalar and real-scalar resonances, respectively, in some benchmark models with dark matter couplings, satisfying the diphoton condition as well as the above current collider bounds.

\subsection{Diphotons from cascade decay}

The ATLAS ECAL is located at $r=1.5\,{\rm meters}$ from the beam and the CMS ECAL lies at $r=1.3\,{\rm meters}$. The cell size of ECAL detectors in CMS and ATLAS is about $\eta=0.0174$ and $0.025$ in pseudo-rapidity, respectively. We also note that the first layer of the ATLAS ECAL ranges between $0.003$ and $0.006$ depending on $\eta$. 
So, if $|\Delta\eta|$ between two photons is smaller than the ECAL cell size, two photons would hit the same ECAL cell so they are identified as a single photon in the ECAL detector \cite{cascade}.

Suppose that diphotons come from the cascade decay of the resonance through light intermediate particles  \cite{cascade}, namely, $X\rightarrow YY\rightarrow 4\gamma$ with $Y\rightarrow \gamma\gamma$. Then, for $m_X\gg m_Y$, the decay length of the $Y$ particle is given by
\be
d=(c\tau_Y)\gamma\approx \frac{1}{\Gamma_Y}\,\frac{E_Y}{m_Y},
\ee
with $\gamma=E_Y/m_Y$. 
On the other hand, the pseudo-rapidity separation between a photon pair coming from the decay of the $Y$ particle is given by
\be
|\Delta \eta|\approx  \frac{2m_Y}{E_Y} \Big(1-\frac{d}{r}\Big).
\ee
For instance, for $d\lesssim r$ and $|\Delta\eta|<0.003$, taking $E_Y=m_X/2$ with $m_X=750\,{\rm GeV}$, we need $m_Y\lesssim 0.5\,{\rm GeV}$.
In this case, two photons coming from the decay of each $Y$ are collimated and are considered as a singlet photon in the detector. 
Then, the resonance production cross section of particle $X$ with cascade decays is
\bea
\sigma(pp\rightarrow X\rightarrow 4\gamma)=\frac{1}{sM_X \Gamma_X}\, K_{gg} C_{gg}\Gamma(X\rightarrow gg) \Gamma(X\rightarrow YY) ({\rm BR}(Y\rightarrow \gamma\gamma))^2.
\eea
We note that the $Y$ couplings to gauge bosons can be small enough as far as the decay length is smaller than the ECAL radius. Thus, the bounds on a light scalar at the LEP or LHC can be evaded.

\begin{figure}
  \begin{center}
   \includegraphics[height=0.50\textwidth]{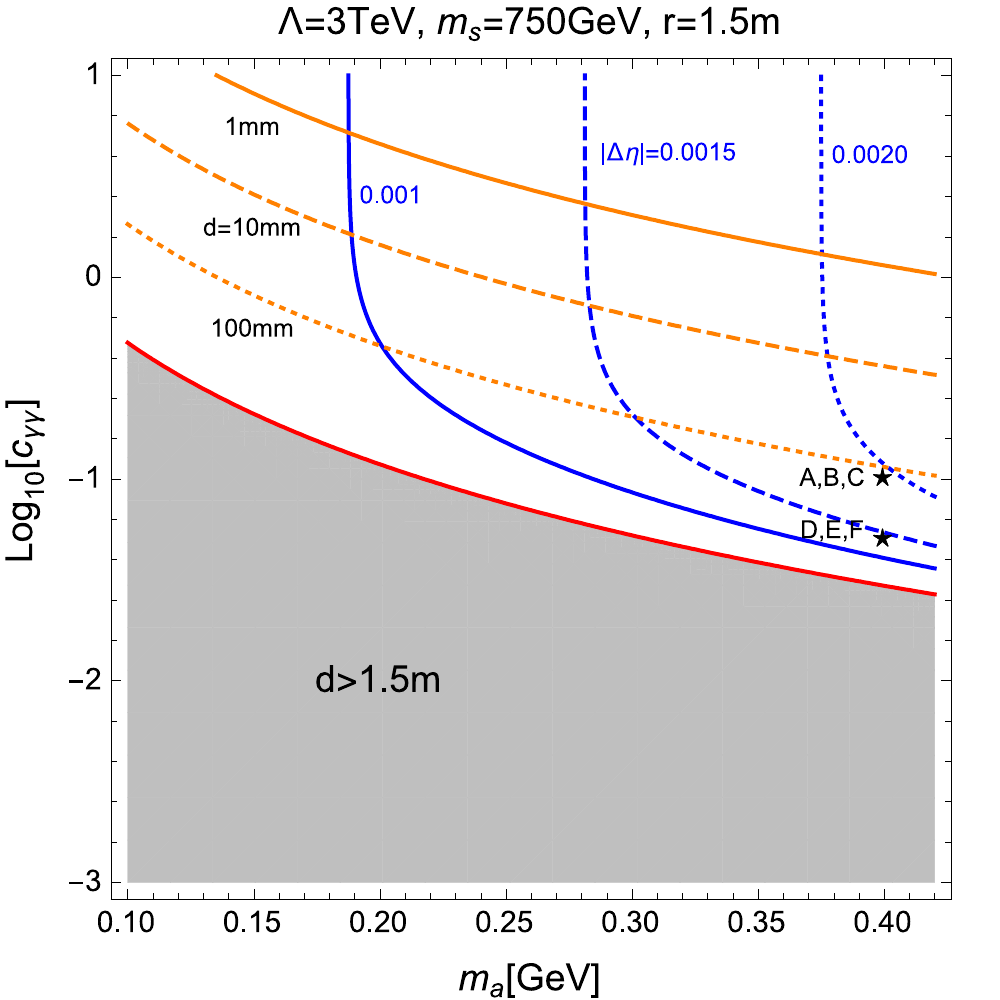}
   \end{center}
  \caption{Contours for separation of two photons in pseudo-rapidity $|\Delta\eta|$ and decay length $d$ in the parameter space of the mass $m_a$ and photon coupling $c_{\gamma\gamma}$ of the light pseudo-scalar.   $\Lambda=3\,{\rm TeV}$ and $m_s=750\,{\rm GeV}$ are taken, while $c_2=0$, $c_1\neq 0$ and $c_3\neq 0$ is taken in the unbroken phase. The radius of ECAL is chosen to $r=1.5\,{\rm m}$. The region with $d>r=1.5\,{\rm m}$ is shown in gray, with boundary $d=1.5\,{\rm m}$ in red line. We have shown benchmark models in star, as will be discussed in Table~\ref{BRs-two}. }
  \label{cascade0}
\end{figure}

In our model, we choose a real scalar resonance of $750\,{\rm GeV}$ and a light pseudo-scalar having mass $m_a$, namely, $X=h_2$ and $Y=a$. In Fig.~\ref{cascade0}, we depict the separation of two photons $|\Delta\eta|$ and the decay length $d$ for the decay of the pseudo-scalar in the parameter space for the mass and photon coupling of the pseudo-scalar. 
For a lot of the parameter space, the pseudo-scalar can decay well within the ECAL radius. 
For instance, for $m_2=750\,{\rm GeV}$ and $m_a=0.4\,{\rm GeV}$,  we need the photon coupling to be $c_{\gamma\gamma}>0.028$ for $d<1.5\,{\rm m}$. Moreover, the region of the parameter space with $c_{\gamma\gamma}\lesssim 0.4$ shown for the cascade decay in Fig.~\ref{cascade0} is consistent with the previous limits from $e^+ e^-\rightarrow 
\gamma^* /Z^{(*)}\rightarrow a\gamma$ with $a\rightarrow 2\gamma$ in LEP \cite{jaeckel}, in particular, at the $Z$-peak \footnote{The hypercharge coupling to the pseudo-scalar can lead to the $Z\gamma$ coupling as well as the $\gamma\gamma$ coupling. }. But, LHC and future colliders such as FCC-ee would be able to probe the photon coupling $c_{\gamma\gamma}$ of order $0.01$ for sub-GeV masses from the same process  \cite{jaeckel}.

\begin{figure}
  \begin{center}
   \includegraphics[height=0.33\textwidth]{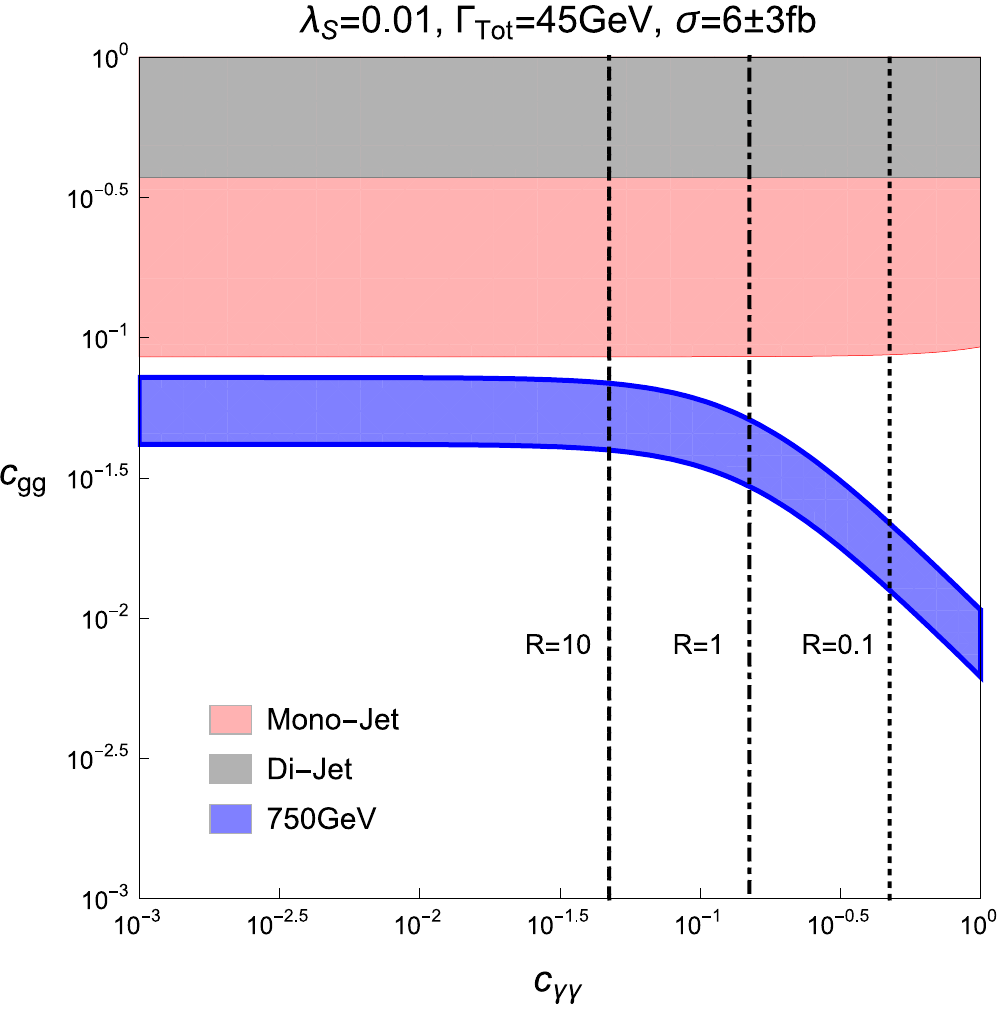}
    \includegraphics[height=0.33\textwidth]{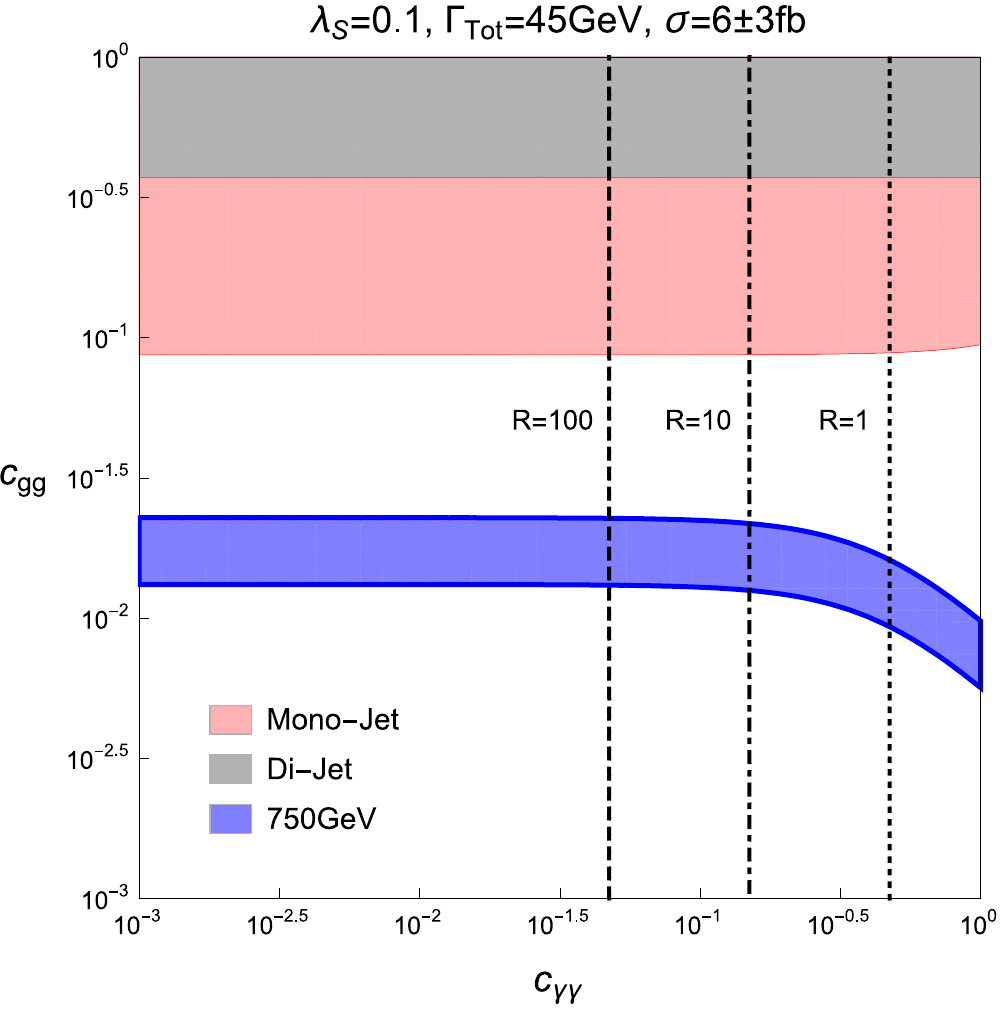} 
    \includegraphics[height=0.33\textwidth]{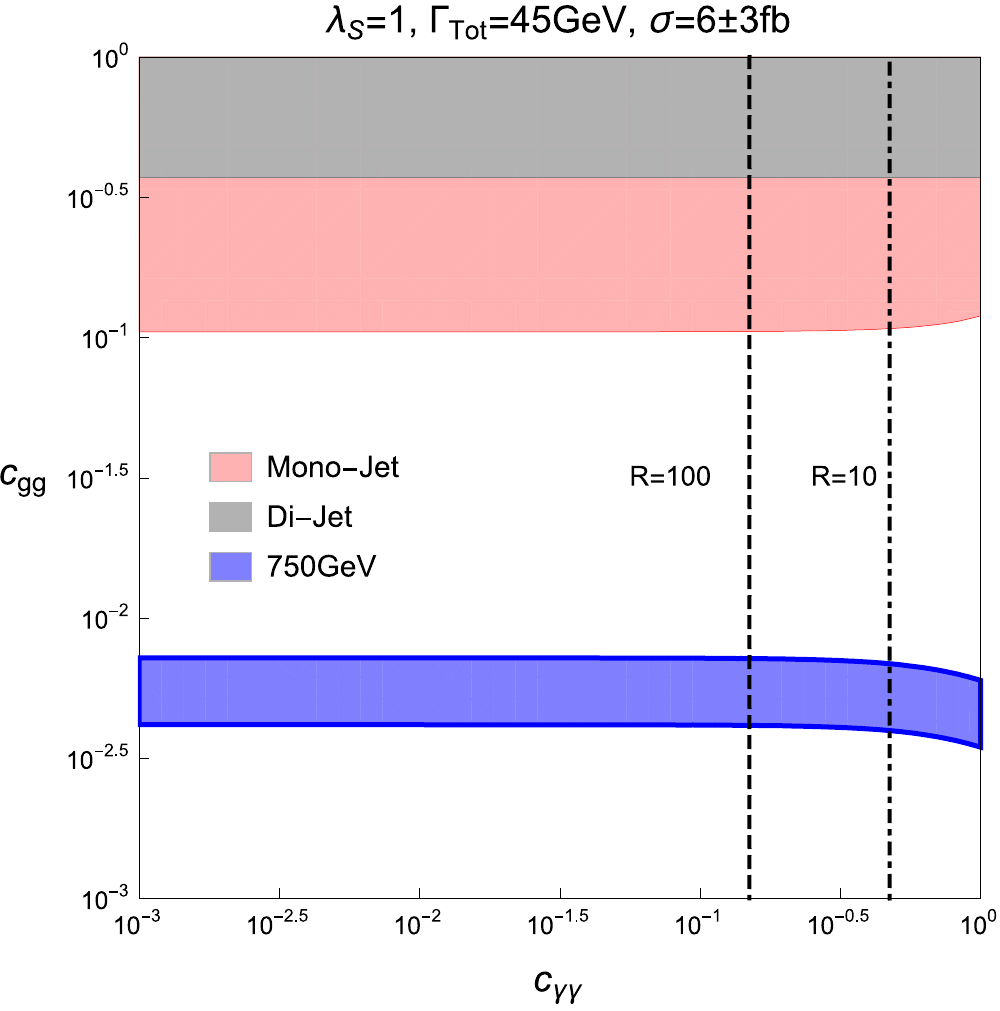}
   \end{center}
  \caption{Parameter space of $c_{\gamma\gamma}$ and $c_{gg}$ for the diphoton resonance, including both direct and cascade decays of the real scalar. We have taken $c_2=0$, $c_1\neq 0$ and $c_3\neq 0$ in the unbroken phase. We have chosen $m_s=750\,{\rm GeV}$ and $m_a=0.4\,{\rm GeV}$. The region explaining the diphoton resonance at $750\,{\rm GeV}$ for $\sigma(pp\rightarrow\gamma\gamma)=6\pm 3\,{\rm fb}$ is shown in blue. Several values of the ratio of cascade to direct decay rates, $R$, are shown in each plot.
   The regions excluded by mono-jet and di-jet limits from LHC $8\,{\rm TeV}$ are shown in pink and gray, respectively. The quartic coupling of the real scalar is chosen to $\lambda_S=0.01, 0.1, 1.0$, from left to right plots. In all the plots, the total decay rate of the real scalar, including the invisible decay mode, is fixed to $\Gamma_{\rm tot}=45\,{\rm GeV}$. }
  \label{cascade}
\end{figure}

With the contribution from the direct decay of the resonance into two photons included in our model, the observed diphoton production cross section leads to 
\bea
\Gamma_2(gg)[\Gamma_2(\gamma\gamma)+ \Gamma_2(aa)({\rm BR}(a\rightarrow\gamma\gamma))^2]=2.8\times 10^{-2}{\rm GeV}^2 \Big(\frac{\sigma(pp\rightarrow \gamma\gamma)}{6\,{\rm fb}} \Big) \Big(\frac{\Gamma_2}{45\,{\rm GeV}}\Big).  \label{diphoton-as}
\eea 
In this case, a small gluon coupling is allowed for a sizable partial decay rate of the real-scalar into a pair of pseudo-scalars, as far as ${\rm BR}(a\rightarrow \gamma\gamma)$ is sizable.  
In Fig.~\ref{cascade}, we show the parameter space for $c_{gg}$ vs $c_{\gamma\gamma}$ by including the cascade decay contribution to the diphoton excess, denoted by the ratio of cascade to direct decay into photons, $R\equiv  \Gamma_2(aa)({\rm BR}(a\rightarrow\gamma\gamma))^2/\Gamma_2(\gamma\gamma)$. For ${\rm BR}(a\rightarrow\gamma\gamma)=1$, we have taken\footnote{For $m_a>3m_\pi$, the pseudo-scalar decays into three pions or mesons, so ${\rm BR}(a\rightarrow\gamma\gamma)$ gets suppressed. In this case, the cascade contribution to diphoton excesses is sub-dominant. Furthermore, the $\eta$ separation between collimated photons becomes larger than $0.003$ for $m_a\gtrsim 0.5\,{\rm GeV}$. } $m_a\lesssim 3m_\pi$, namely, $m_a=0.4\,{\rm GeV}$. Keeping the total width of the real-scalar resonance to $\Gamma_{s,\rm tot}=45\,{\rm GeV}$, we vary the singlet quartic coupling $\lambda_S=0.01, 0.1, 1$, from left to right figures in Fig.~\ref{cascade}, and show that the parameter space with cascade decay dominance increases, being compatible with mono-jet and di-jet bounds.  In the case with a sizable cascade decay, the gluon coupling is more or less fixed to a small value while there is a little dependence on the photon coupling as far as the photon-jet contribution is dominant.

When the diphoton resonance is dominated by the cascade decay, the condition on the effective couplings of the resonance becomes
\bea
|d_{gg}| \sqrt{\lambda_S}= 0.071\Big(\frac{\sigma(pp\rightarrow \gamma\gamma)}{6\,{\rm fb}} \Big)^{1/2} \Big(\frac{\Lambda}{3\,{\rm TeV}}\Big)^2 \Big(\frac{\Gamma_2}{45\,{\rm GeV}} \Big)^{1/2}\Big(\frac{1/9}{{\rm BR}(a\rightarrow \gamma\gamma)}\Big)
\eea
where use is made of $m_2=\sqrt{2\lambda_S} v_s$ and $m_a\ll m_s/2$.  
In the decoupled vector-like fermions with the same global charges as dark matter, we get $d_{gg}=\frac{4}{3} c_{gg}$, so the above condition becomes
\bea
|c_{gg}| \sqrt{\lambda_S}= 0.053\Big(\frac{\sigma(pp\rightarrow \gamma\gamma)}{6\,{\rm fb}} \Big)^{1/2} \Big(\frac{\Lambda}{3\,{\rm TeV}}\Big)^2 \Big(\frac{\Gamma_2}{45\,{\rm GeV}} \Big)^{1/2}\Big(\frac{1/9}{{\rm BR}(a\rightarrow \gamma\gamma)}\Big)
\eea

\begin{table}[ht]
\centering
\small
\begin{tabular}{|c||c|c|c|c|}
\hline 
Model & ${\rm BR}_s(\gamma\gamma)$ & ${\rm BR}_s(gg)$ & ${\rm BR}_s(Z\gamma)$ & ${\rm BR}_s(ZZ)$  \\ [0.7ex] 
\hline 
A & $6.59\times 10^{-3}$ & $3.99\times 10^{-4}$ & $3.81\times 10^{-3}$ & $5.42\times 10^{-4}$ \\ [0.7ex]
\hline
B & $4.14\times 10^{-2}$ & $3.99\times 10^{-4}$ & $2.39\times 10^{-2}$ & $3.40\times 10^{-3}$  \\ [0.7ex]
\hline
C & $1.68\times 10^{-3}$ & $4.15\times 10^{-5}$ & $9.71\times 10^{-4}$ & $1.38\times 10^{-4}$  \\ [0.7ex]
\hline
D & $1.99\times 10^{-3}$ & $4.12\times 10^{-4}$ & $1.15\times 10^{-3}$ & $1.64\times 10^{-4}$ \\ [0.7ex]
\hline
E & $1.09\times 10^{-2}$ & $4.12\times 10^{-4}$ & $6.31\times 10^{-3}$ & $8.97\times 10^{-4}$  \\ [0.7ex]
\hline
F & $4.38\times 10^{-4}$ & $4.17\times 10^{-5}$ & $2.53\times 10^{-4}$ & $3.60\times 10^{-5}$ \\ [0.7ex]
\hline
\end{tabular}
\vskip 0.3cm
\begin{tabular}{|c||c|c|c|c|}
\hline 
Model & ${\rm BR}_s(\chi\bar{\chi})$ & ${\rm BR}_s(aa)$ & $\Gamma_{s,{\rm tot}}$ & $\Gamma_{a,{\rm tot}}$ \\ [0.7ex]
\hline
A &  $0.840$ & $0.148$ & $10.1$ & $5.66\times 10^{-12}$ \\ [0.7ex]
\hline
B & - & $0.931$ & $1.60$ & $5.66\times 10^{-12}$\\ [0.7ex]
\hline 
C  & $0.620$ & $0.378$ & $39.5$ & $5.66\times 10^{-12}$\\ [0.7ex]
\hline
D  & $0.817$ & $0.167$ & $8.32$ & $1.42\times 10^{-12}$ \\ [0.7ex]
\hline
E & - & $0.982$ & $1.52$ & $1.42\times 10^{-12}$\\ [0.7ex]
\hline
F & $0.605$ & $0.394$ & $37.9$ & $1.42\times 10^{-12} $ \\ [0.7ex]
\hline
\end{tabular}
\caption{Decay branching fractions and total decay rates (in units of GeV) of real-scalar and pseudo-scalar, when the former is the $750\,{\rm GeV}$ resonance.
Benchmark models with $(c_{\gamma\gamma }, m_\chi, \lambda_\chi, \lambda_s, c_{gg})$ are 
Model A: $(0.1,320\,{\rm GeV},2.0,0.1,4.29\times 10^{-3})$,
Model B: $(0.1,950\,{\rm GeV},1.0,0.1,3.47\times 10^{-3})$,
Model C: $(0.1,190\,{\rm GeV},1.6,1.0,5.56\times 10^{-3})$,
Model D: $(0.05,328\,{\rm GeV},2.0,0.1,3.44\times 10^{-3})$,
Model E: $(0.05,920\,{\rm GeV},1.0,0.1,3.44\times 10^{-3})$,
Model F: $(0.05,220\,{\rm GeV},1.7,1.0,3.43\times 10^{-3})$.
We have taken $c_2=0$, $c_1\neq 0$ and $c_3\neq 0$ in the unbroken phase. For all models, we have imposed $\sigma(pp\rightarrow\gamma\gamma)=6\,{\rm fb}$ for $m_s=750\,{\rm GeV}$ and $m_a=0.4\,{\rm GeV}$ while the current collider bounds and $d<1.5\,{\rm m}$ are fulfilled. These benchmark models will be used for dark matter discussion in Table~\ref{ann-as} in Section 6.}
\label{BRs-two}
\end{table}

The mono-jet bound from CMS $8\,{\rm TeV}$ for the case with cascade decay is given as follows,
\bea
\frac{\Gamma_2({\chi{\bar\chi}})}{\Gamma_2(\gamma\gamma)+\Gamma_2(aa)({\rm BR}(a\rightarrow \gamma\gamma))^2} < 667 \Big(\frac{r}{5}\Big) \left(\frac{6\,{\rm fb}}{\sigma(pp\rightarrow \gamma\gamma)} \right).
\eea
When the cascade decay is dominant, for $r=4.7$ and $\sigma(pp\rightarrow\gamma\gamma)=6\,{\rm fb}$, we get the mono-jet bound on the dark matter coupling to the resonance as
\bea
\frac{\lambda^2_\chi m^2_2}{2\lambda^2_S v^2_s} \lesssim 9.80  \Big(\frac{{\rm BR}(a\rightarrow \gamma\gamma)}{1/9} \Big)^2,
\eea
which becomes, for $\Lambda=3\,{\rm TeV}$ and $m_2=750\,{\rm GeV}$,
\be
|\lambda_\chi| \lesssim 3.13\sqrt{\lambda_S} \Big(\frac{{\rm BR}(a\rightarrow \gamma\gamma)}{1/9} \Big).
\ee
Here, use is made of $m_2=\sqrt{2\lambda_S}\, v_s$ in the limit of a vanishing Higgs mixing angle. 

On the other hand, the di-jet bound at  LHC 8 TeV leads to 
\bea
\frac{\Gamma_2(gg)+\Gamma_2(aa)({\rm BR}(a\rightarrow gg))^2}{\Gamma_2(\gamma\gamma)+\Gamma_2(aa)({\rm BR}(a\rightarrow \gamma\gamma))^2} \lesssim 2083  \,\Big(\frac{r}{5}\Big) \Big(\frac{6\,{\rm fb}}{\sigma(pp\rightarrow\gamma\gamma )}\Big).
\eea
When the cascade decay is dominant, for $r=4.7$ and $\sigma(pp\rightarrow\gamma\gamma)=6\,{\rm fb}$, the gluon coupling to the real scalar resonance is constrained by
\bea
|c_{gg}|\lesssim 2.35|c_{\gamma\gamma}|.
\eea
Therefore, the gluon coupling is much more constrained, as compared to the case with direct decay where $|c_{gg}|\lesssim 15.6|c_{\gamma\gamma}|$ is obtained from the di-jet bound. 

On the other hand, if the pseudo-scalar is lighter than $3m_\pi\sim 420\,{\rm MeV}$, we get ${\rm BR}(a\rightarrow \gamma\gamma)=1$. In this case, there is no extra di-jet from the cascade decays.
Then, for $r=4.7$, $\sigma(pp\rightarrow\gamma\gamma)=6\,{\rm fb}$, $\Lambda=3\,{\rm TeV}$ and $m_2=750\,{\rm GeV}$, the dijet bound  leads to 
$|d_{gg}|\lesssim 32 \sqrt{\lambda_S}$.

In the presence of cascade decay of the real scalar, in Table~\ref{BRs-two}, we show the branching fractions and total decay rate of the pseudo-scalar and real-scalar, respectively, in some benchmark models with dark matter couplings, that satisfy the diphoton condition as well as the above current collider bounds.  Here, we have set the scalar masses to $m_s=750\,{\rm GeV}$ and $m_a=0.4\,{\rm GeV}$ below the pion threshold such that ${\rm BR}(a\rightarrow \gamma\gamma)=1$, and the effective gauge couplings are taken to $d_i=\frac{4}{3}c_i (i=1,3)$ and $c_2=d_2=0$. 
The photon couplings in all the models are within the reach of the future colliders.

\section{Dark matter with pseudo-scalar resonance}

In this section, we interpret the diphoton resonance by the direct decay of the pseudo-scalar in our model, focusing on the pseudo-scalar coupling to a Dirac singlet fermion dark matter.  The  DM relic density condition, the constraints from indirect detection for dark matter and the mono-jet  limits are superimposed.

\subsection{Dark matter annihilation}

\begin{figure}
  \begin{center}
   \includegraphics[height=0.15\textwidth]{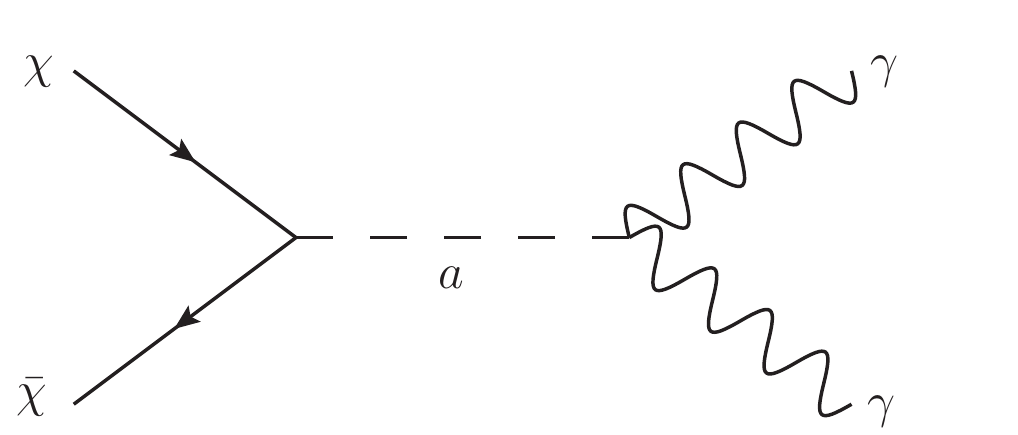}
    \includegraphics[height=0.15\textwidth]{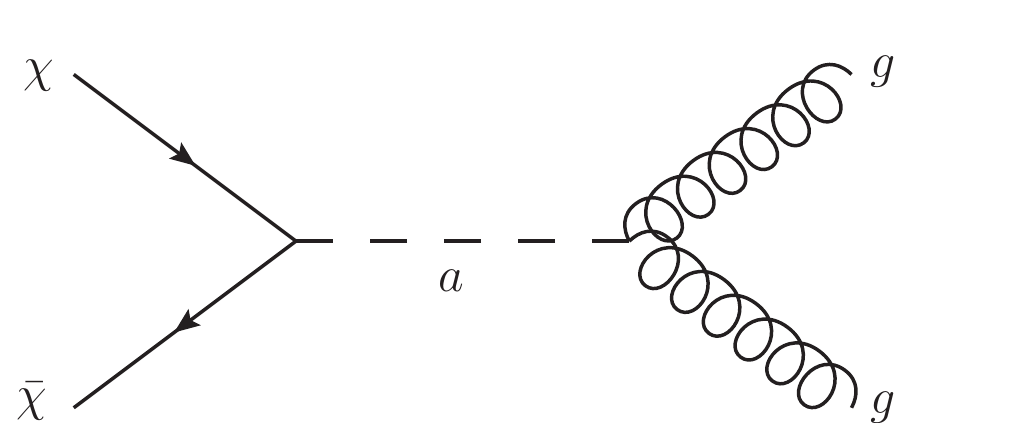}
     \includegraphics[height=0.15\textwidth]{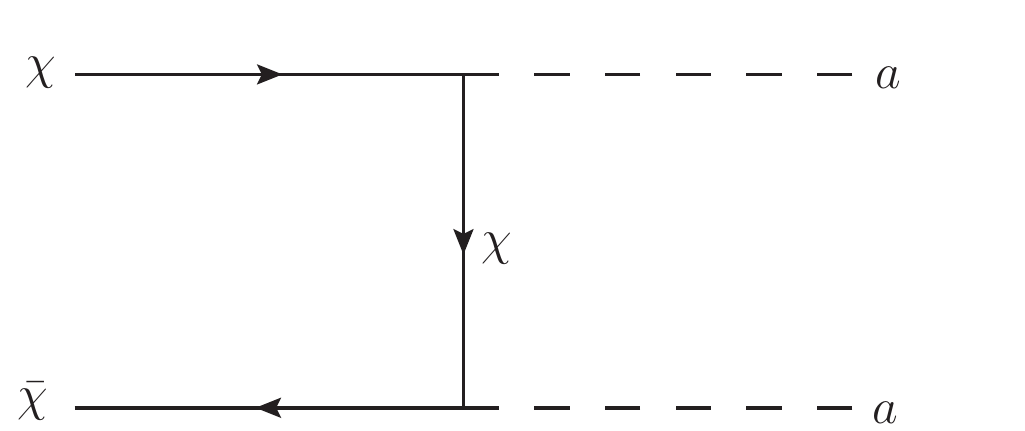}
   \end{center}
  \caption{Feynman diagrams for dark matter annihilation with pseudo-scalar resonance. }
  \label{a-diagrams}
\end{figure}

When the real scalar is heavy, we can consider the interactions of the pseudo-scalar field only in the Lagrangian (\ref{effaction}).
Then, the pseudo-scalar can play a role of mediator between dark matter and the SM \cite{lpp,MET,ibarra}. 
When vector-like fermions are sufficiently heavier than dark matter,  we can use the effective interactions for pseudo-scalar resonance in the process of dark matter annihilation as shown in Fig.~\ref{a-diagrams}.
In this case, the total annihilation cross section of dark matter is given by $(\sigma v_{\rm rel})_a=\sum_i (\sigma v_{\rm rel})_{a,i}+(\sigma v_{\rm rel})_{aa}$ with partial annihilation cross sections into a pair of SM gauge bosons being
\bea
(\sigma v_{\rm rel})_{a,gg}&=& \frac{16\lambda^2_\chi c^2_{gg}}{\pi \Lambda^2}\,\frac{m^4_\chi}{(4m^2_\chi-m^2_a)^2+\Gamma^2_a m^2_a}, \\
(\sigma v_{\rm rel})_{a,\gamma\gamma}&=& \frac{2\lambda^2_\chi c^2_{\gamma\gamma}}{\pi\Lambda^2}\,\frac{m^4_\chi}{(4m^2_\chi-m^2_a)^2+\Gamma^2_a m^2_a}, \\
(\sigma v_{\rm rel})_{a,Z\gamma} &=&  
 \frac{\lambda^2_\chi c^2_{Z\gamma}}{\pi\Lambda^2}\,\frac{m^4_\chi}{(4m^2_\chi-m^2_a)^2+\Gamma^2_a m^2_a}\,\left( 1-\frac{m^2_Z}{4m^2_\chi} \right)^3, \\
 ( \sigma v_{\rm rel})_{a,ZZ} &=&  \frac{2\lambda^2_\chi c^2_{ZZ}}{\pi\Lambda^2} \,\frac{m^4_\chi}{(4m^2_\chi-m^2_a)^2+\Gamma^2_a m^2_a}\,\left(1-\frac{m^2_Z}{m^2_\chi} \right)^{3/2}, \\
 ( \sigma v_{\rm rel})_{a,WW}&=&  \frac{\lambda^2_\chi c^2_{WW}}{\pi\Lambda^2}\,\frac{m^4_\chi}{(4m^2_\chi-m^2_a)^2+\Gamma^2_a m^2_a} \left(1-\frac{m^2_W}{m^2_\chi} \right)^{3/2}
 \eea
where 
\bea
c_{gg}&=& c_3, \\
c_{\gamma\gamma}&=&c_1 \cos^2\theta_W +c_2 \sin^2\theta_W, \\
c_{Z\gamma}&=& (c_2-c_1) \sin(2\theta_W), \\
c_{ZZ} &=&  c_1\sin^2\theta_W+  c_2 \cos^2\theta_W, \\
c_{WW} &=&2c_2.
\eea
We note that all the gauge boson channels are s-wave.

For $m_\chi>m_a$, dark matter can annihilate into a pair of pseudo-scalars.  In the limit of non-relativistic dark matter, the corresponding annihilation cross sections for the $aa$ channel becomes
\bea
(\sigma v_{\rm rel})_{aa}&=& \frac{\lambda^2_\chi}{96\pi} \bigg[ \frac{\lambda^2_\chi m^6_\chi}{(m^2_a-2m^2_\chi)^4} \Big(1-\frac{m^2_a}{m^2_\chi}\Big)^2+\frac{3\lambda^2_S v^2_s}{2[(4m^2_\chi-m^2_s)^2+\Gamma^2_s m^2_s]} \nonumber \\
&& +\frac{\sqrt{2} \lambda_\chi\lambda_S v_s m^3_\chi}{(m^2_a- 2m^2_\chi)^2}\cdot \frac{4m^2_\chi-m^2_s}{(4m^2_\chi-m^2_s)^2+\Gamma^2_s m^2_s}\Big(1-\frac{m^2_a}{m^2_\chi} \Big) \bigg]
\Big(1-\frac{m^2_a}{m^2_\chi}\Big)^{1/2}\, v^2_{\rm rel}\,.
\eea
Here, we have also included the real-scalar contribution to the $aa$ channel, for a later use with real and pseudo-scalars in the effective field theory in Section 6.  
Thus, the $aa$ channel turns out to be $p$-wave suppressed, so they are not relevant for indirect detection at present. However, the $aa$ channel, if open kinematically, still contributes to the thermal cross section at freeze-out.

\subsection{Bounds from indirect detections}

\begin{figure}
  \begin{center}
   \includegraphics[height=0.40\textwidth]{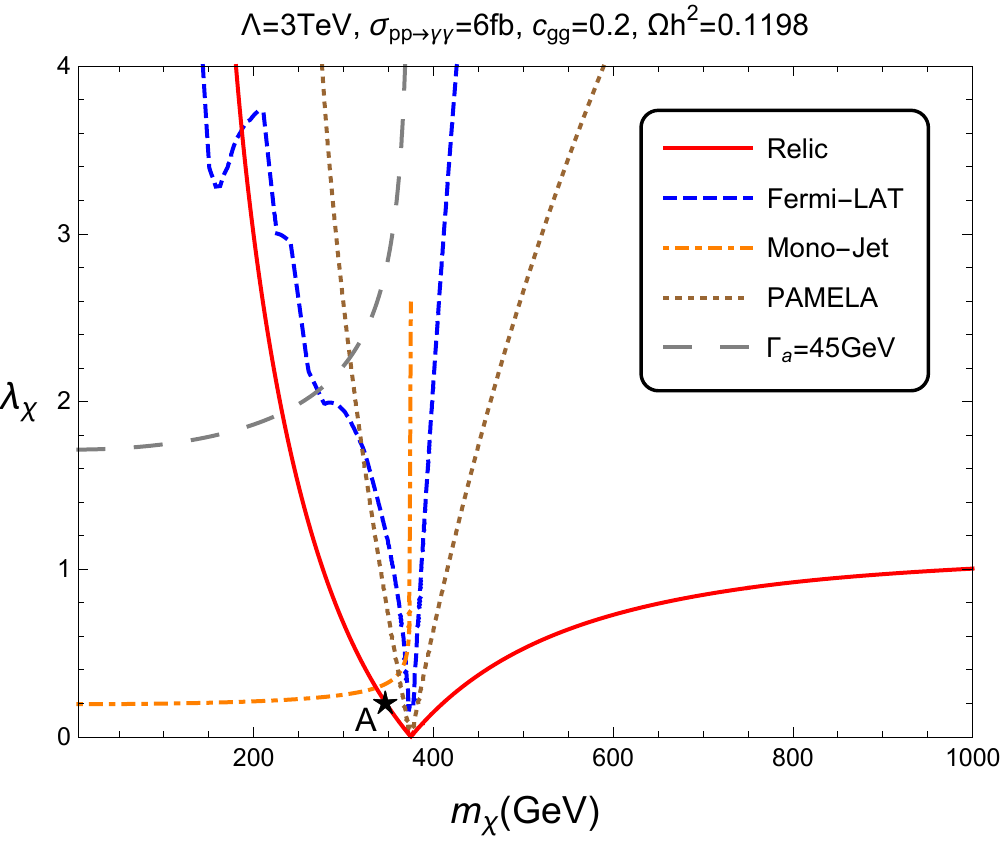}
    \includegraphics[height=0.40\textwidth]{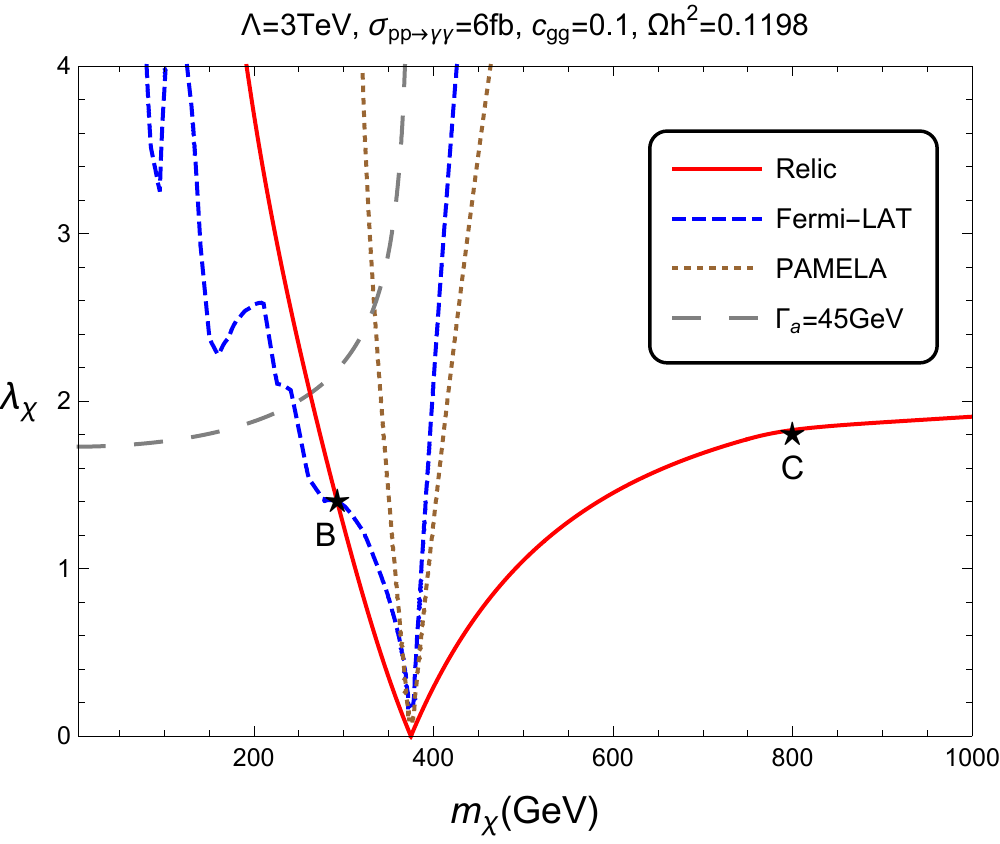} 
   \end{center}
  \caption{Parameter space of $m_\chi$ and $\lambda_\chi$ with pseudo-scalar mediator, satisfying the relic density in red lines. The region explaining the diphoton resonance at $750\,{\rm GeV}$ for $\sigma(pp\rightarrow\gamma\gamma)=6\,{\rm fb}$ is imposed. $c_{gg}=0.2$ and $c_{gg}=0.1$ are chosen on left and right, while $c_2=0$ in both plots.
   The mono-jet  limit from LHC $8\,{\rm TeV}$ are shown in orange dot-dashed line.
   The region above blue dashed line and brown dotted line are excluded by the bound from Fermi-LAT line search (R16 with Einasto profile) and the antiproton bound from PAMELA. 
   The line for $\Gamma_a=45\,{\rm GeV}$ is also shown in dashed gray. Benchmark models  A (B and C) are shown in star on the left (right) plot, taken in Table~\ref{BRa}. }
  \label{relic1}
\end{figure}

The cosmic ray flux stemming from the annihilation of a Dirac fermion dark matter into $f$ final states (such as $\gamma$-ray, $e^+$, $p^+$, $\nu$, etc) is given by
\bea
\frac{d\Phi_f}{dE_f}= \frac{1}{16\pi m^2_\chi} \,\langle\sigma v\rangle_f\,\frac{d N_f}{dE_f}\, J
\eea
where $\frac{dN_f}{dE}$ is the differential cosmic-ray yield per annihilation and the J-factor is the line-of-sight integral through the dark matter distribution integrated over the solid angle $\Delta\Omega$, given by
\be
J=\frac{1}{\Delta \Omega}\int_{\Delta\Omega} d\Omega \int_{\rm l.o.s.} ds\, \rho^2_\chi. 
\ee

On the other hand, the loop-induced interactions of the pseudo-scalar to gluons lead to the effective interactions between dark matter and gluon fields. Thus, gluons inside nucleons can scatter off with dark matter, leading to recoil energy signals in underground experiments. But, the current dark matter experiments are not sensitive enough to detect the signals.
Therefore, henceforth we focus on the indirect detection.
 
Dark matter annihilation channels,  $\chi{\bar\chi}\rightarrow a\rightarrow \gamma\gamma, Z\gamma$, are $s$-wave and they lead to monochromatic photons at $E_\gamma=m_\chi$ and $E_\gamma=m_\chi \Big(1-\frac{4m^2_Z}{m^2_\chi} \Big)$, respectively.  Those channels can be constrained by Fermi-LAT \cite{fermilat2} and HESS \cite{hess2013} line searches from the galactic center.  

Annihilation channels of dark matter into $WW, ZZ, gg$ lead to continuum photons from bremstrahlung or decay and they are constrained by Fermi-diffuse gamma-ray searches from dwarf galaxies \cite{dwarfgalaxy}. 
Moreover, dark matter annihilation into a pair of gluons can be constrained by anti-proton data from PAMELA and AMS-02 \cite{antiproton}.

\begin{table}[ht]
\centering
\small
\begin{tabular}{|c||c|c|c|c|c|c|}
\hline 
Model & $\langle\sigma v_{\rm rel}\rangle_{a,\gamma\gamma}$ & $\langle\sigma v_{\rm rel}\rangle_{a,gg}$ & $\langle\sigma v_{\rm rel}\rangle_{a,Z\gamma}$ & $\langle\sigma v_{\rm rel}\rangle_{a,ZZ}$ & $\langle\sigma v_{\rm rel}\rangle_{aa}$ & $\Omega_\chi h^2$   \\ [0.7ex] 
\hline 
A & $1.48\times 10^{-29}$ & $2.34\times 10^{-26}$ & $8.50\times 10^{-30}$ & $1.20\times 10^{-30}$ & $-$ & $0.122$ \\ [0.7ex]
\hline
B & $2.62\times 10^{-27}$ & $1.98\times 10^{-26}$ & $1.47\times 10^{-27}$ & $2.02\times 10^{-28}$ & $-$ & $0.120$ \\ [0.7ex]
\hline
C & $4.68\times 10^{-29}$ & $2.20\times 10^{-26}$ & $2.80\times 10^{-29}$ & $4.13\times 10^{-30}$ & $6.15\times 10^{-28}$ & $0.124$ \\ [0.7ex]
\hline
\end{tabular}
\caption{Averaged annihilation cross sections (in units of ${\rm cm^3/s}$) at present and relic density for dark matter with pseudo-scalar, except that that the one for the $aa$ channel is given at freeze-out. The benchmark models are the same as in Table~\ref{BRa} and Fig.~\ref{relic1}.
All the constraints from the current collider and cosmic data are satisfied. 
}
\label{ann-pseudo}
\end{table}

In Fig.~\ref{relic1}, we show the parameter space of dark matter mass $m_\chi$ and coupling $\lambda_\chi$ in the model with pseudo-scalar resonance where the condition for diphoton excesses is satisfied. Depending on the value of the gluon coupling $c_{gg}=0.2 (0.1)$ on left (right) plots, respectively, with the photon coupling $c_{\gamma\gamma}$ being determined by the diphoton condition (\ref{diphoton-a}), we imposed the current bounds from mono-jet searches as well as the indirect detections.  

In the former case with $c_{gg}=0.2$, the mono-jet bound is quite constraining below resonance, so only the region with small dark matter coupling near resonance survives, while the antiproton bound from PAMELA reaches closely to the region saturating the relic density and the bound from other cosmic data such as Fermi-LAT are not strong. 
In the latter case with $c_{gg}=0.1$, there is no mono-jet bound, but the bound from Fermi-LAT line search constrain most strongly the region with small dark matter masses below resonance, allowing only the small region near resonance. Therefore, the mono-jet and Fermi-LAT line searches are complementary to constraining the light dark matter.  On the other hand, the region above resonance is not constrained in the region where the relic density is saturated.

In Table~\ref{ann-pseudo}, we show the averaged annihilation cross sections at present (except the one for the $aa$ channel, which is given at freeze-out) and the relic density for dark matter with pseudo-scalar mediator in some benchmark models considered in Table~\ref{BRa}, satisfying the diphoton condition as well as the above current collider bounds.  
Model A (B and C) belongs to the left (right) plot in Fig. ~\ref{relic1}.
These models satisfy the current bounds from various indirect detection experiments discussed above.

\section{Dark matter with real-scalar resonance}

In this section, we consider the real-scalar resonance for the diphoton excess and discuss the interplay with indirect detection of dark matter and mono-jet searches, similarly to the case with pseudo-scalar case.

\begin{figure}
  \begin{center}
   \includegraphics[height=0.15\textwidth]{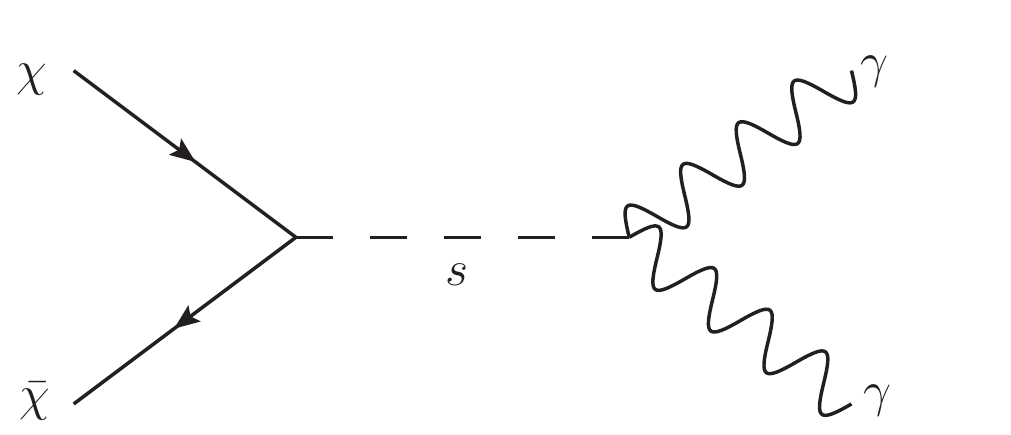}
    \includegraphics[height=0.15\textwidth]{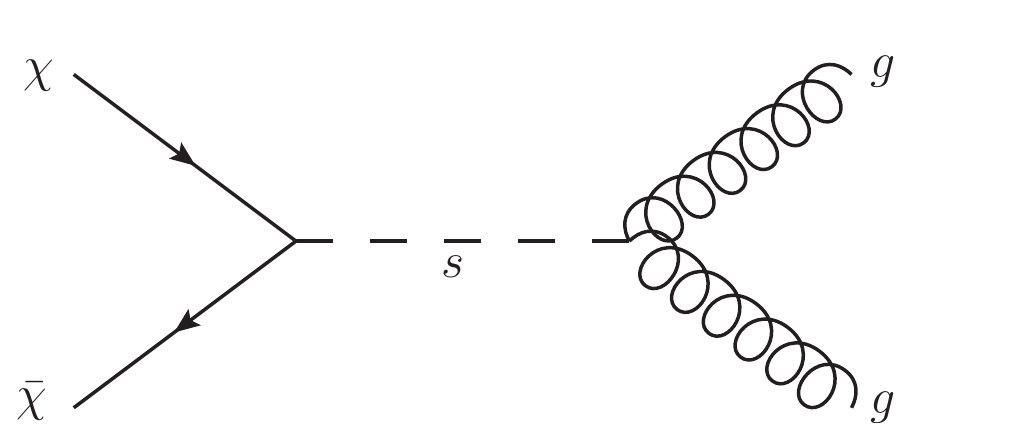}
     \includegraphics[height=0.15\textwidth]{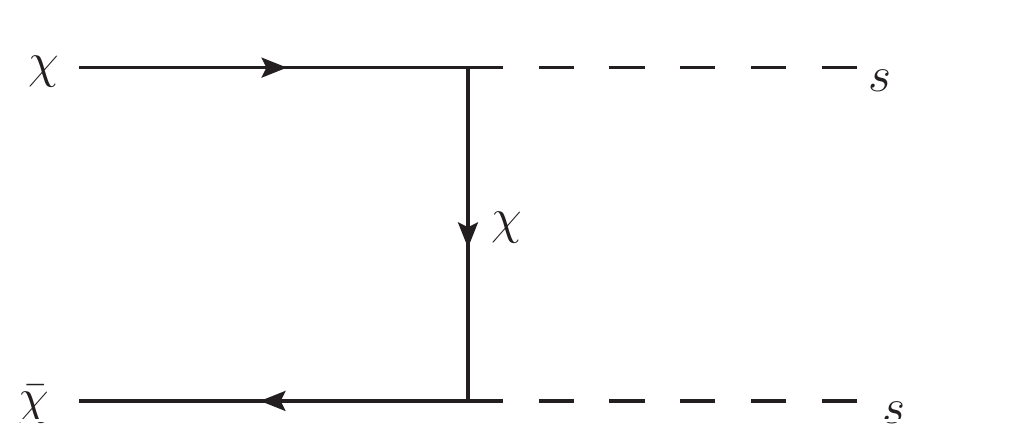}
       \includegraphics[height=0.15\textwidth]{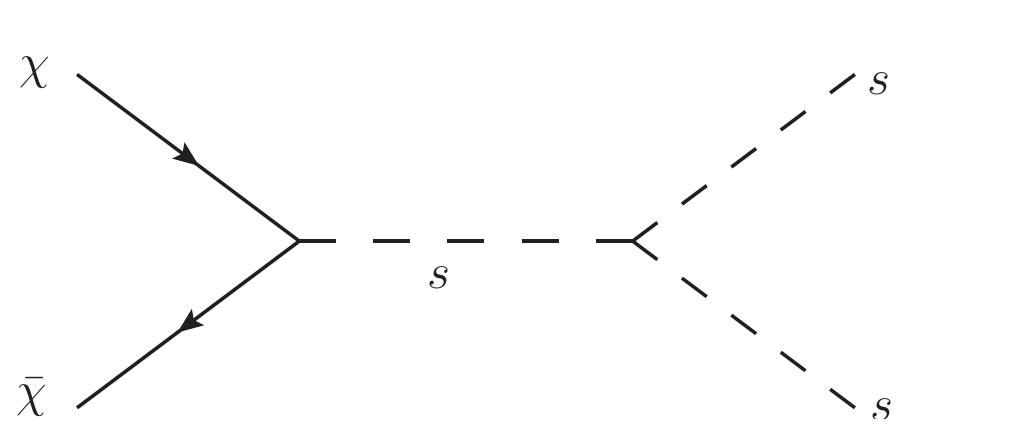}
   \end{center}
  \caption{Feynman diagrams for dark matter annihilation with real-scalar resonance.  }
  \label{s-diagrams}
\end{figure}

When the real scalar is light enough, it can contribute to the DM annihilation through s-channels and/or t-channels as shown in Fig.~\ref{s-diagrams}. 
Taking vector-like fermions in loops to be sufficiently heavier than dark matter, we obtain the total annihilation cross section of dark matter in terms of effective interactions for real-scalar resonance by $(\sigma v_{\rm rel})_s=\sum_i (\sigma v_{\rm rel})_{s,i} +(\sigma v_{\rm rel})_{ss}$ with partial annihilation cross sections into a pair of SM gauge bosons  being given by 
\bea
(\sigma v_{\rm rel})_{s,gg}&=& \frac{4\lambda^2_\chi d^2_{gg}}{\pi \Lambda^2}\,\frac{m^4_\chi}{(4m^2_\chi-m^2_2)^2+\Gamma^2_2 m^2_2}\,v^2_{\rm rel}, \\
(\sigma v_{\rm rel})_{s,\gamma\gamma}&=& \frac{\lambda^2_\chi d^2_{\gamma\gamma}}{2\pi\Lambda^2}\,\frac{m^4_\chi}{(4m^2_\chi-m^2_2)^2+\Gamma^2_2 m^2_2}\,v^2_{\rm rel}, \\
(\sigma v_{\rm rel})_{s,Z\gamma} &=&  
 \frac{\lambda^2_\chi d^2_{Z\gamma}}{4\pi\Lambda^2}\,\frac{m^4_\chi}{(4m^2_\chi-m^2_2)^2+\Gamma^2_2 m^2_2}\,\left( 1-\frac{m^2_Z}{4m^2_\chi} \right)^3 v^2_{\rm rel}, \\
 ( \sigma v_{\rm rel})_{s,ZZ} &=&  \frac{\lambda^2_\chi d^2_{ZZ}}{2\pi\Lambda^2} \,\frac{m^4_\chi v^2_{\rm rel}}{(4m^2_\chi-m^2_2)^2+\Gamma^2_2 m^2_2}\,\Big(1-\frac{m^2_Z}{m^2_\chi}+\frac{3m^4_Z}{8m^4_\chi}\Big)\left(1-\frac{m^2_Z}{m^2_\chi} \right)^{1/2} , \\
 ( \sigma v_{\rm rel})_{s,WW}&=&  \frac{\lambda^2_\chi d^2_{WW}}{4\pi\Lambda^2}\,\frac{m^4_\chi v^2_{\rm rel}}{(4m^2_\chi-m^2_2)^2+\Gamma^2_2 m^2_2}\, \Big(1-\frac{m^2_W}{m^2_\chi}+\frac{3m^4_W}{8m^4_\chi}\Big)\left(1-\frac{m^2_W}{m^2_\chi} \right)^{1/2}.
 \eea
We note that all the gauge boson channels are $p$-wave suppressed.

\begin{figure}
  \begin{center}
   \includegraphics[height=0.40\textwidth]{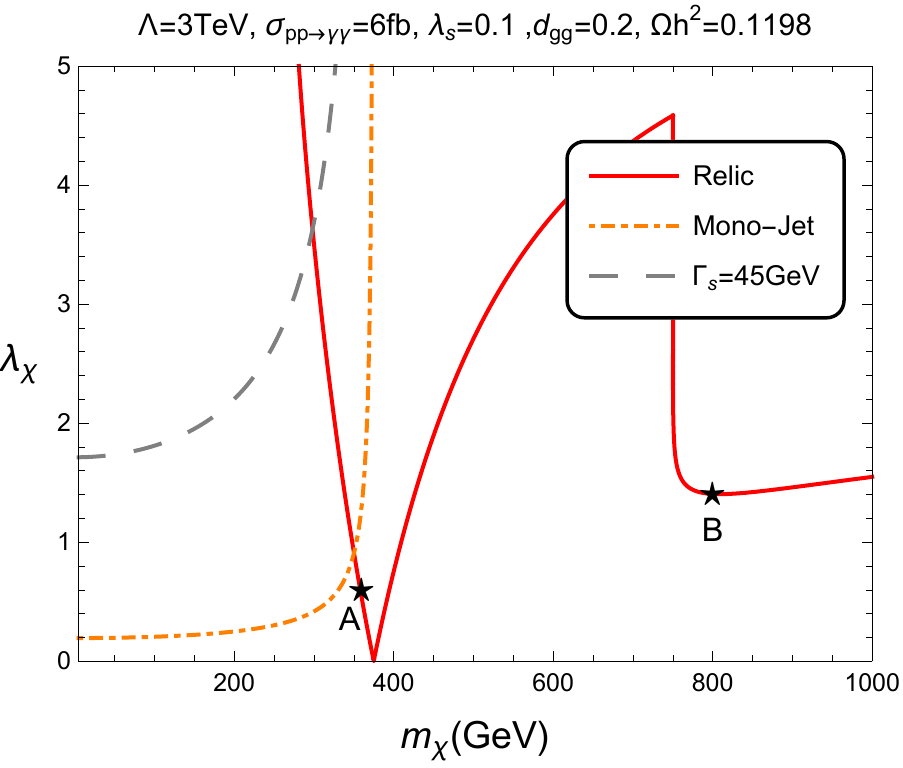}
    \includegraphics[height=0.40\textwidth]{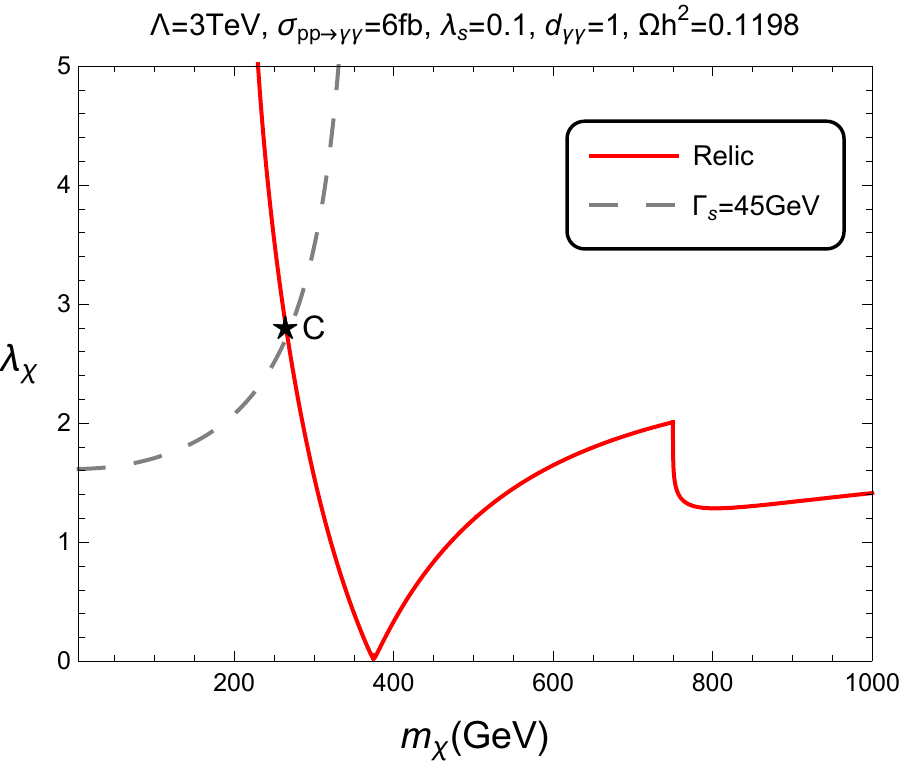} 
   \end{center}
  \caption{Parameter space of $m_\chi$ and $\lambda_\chi$ with real-scalar mediator, satisfying the relic density in red lines. The region explaining the diphoton resonance at $750\,{\rm GeV}$ for $\sigma(pp\rightarrow\gamma\gamma)=6\,{\rm fb}$ is imposed. $d_{gg}=0.2$ and $d_{\gamma\gamma}=1.0$ are chosen on left and right, while $c_2=0$ in both plots.
   The mono-jet bound rom LHC $8\,{\rm TeV}$ is shown in orange dot-dashed line. The line for $\Gamma_s=45\,{\rm GeV}$ is also shown in dashed gray. Benchmark models A and B (C) are shown in star on the left (right) plot, taken in Table~\ref{BRs}.}
  \label{relic2}
\end{figure}

For $m_\chi>m_s$, dark matter can annihilate into a pair of real scalars.  In the limit of non-relativistic dark matter, the corresponding annihilation cross section for the $ss$ channel become  
\bea
(\sigma v_{\rm rel})_{ss}&=&  \frac{\lambda^2_\chi}{32\pi}\bigg[ \frac{\lambda^2_\chi m^2_\chi(2(m^2_s-2m^2_\chi)^2+m^4_\chi)}{3(m^2_s-2m^2_\chi)^4}+\frac{9\lambda^2_S v^2_s }{2[(4m^2_\chi-m^2_s)^2+\Gamma^2_s m^2_s]} \nonumber \\
&&+\frac{6\sqrt{2}\lambda_\chi\lambda_S v_s m_\chi}{m^2_s-2m^2_\chi}\frac{4m^2_\chi-m^2_s}{(4m^2_\chi-m^2_s)^2+\Gamma^2_s m^2_s} \bigg]  \Big(1-\frac{m^2_s}{m^2_\chi}\Big)^{1/2}\, v^2_{\rm rel}\,,
\eea
Thus, the $ss$ channel turns out to be $p$-wave suppressed, but it can contribute to the thermal cross section at freeze-out.  

In Fig.~\ref{relic2}, we show the parameter space of dark matter mass $m_\chi$ and coupling $\lambda_\chi$ in the model with real-scalar resonance where the condition for diphoton excesses is satisfied. Depending on whether the gluon or photon coupling is dominant, namely, $d_{gg}=0.2 (d_{\gamma\gamma}=1.0)$ on left (right) plots, respectively, with the photon or gluon coupling ($d_{\gamma\gamma}$ or $d_{gg}$) being determined by the diphoton condition (\ref{diphoton-a}), we imposed the current bounds from mono-jet searches. In the former case with a large gluon coupling, the mono-jet bound is still strong below resonance, as in the case with pseudo-scalar resonance.
But, in the latter case with a large photon coupling and accordingly a small gluon coupling due to the diphoton condition, there is no bound from mono-jet searches. 
As the dark matter annihilation into a pair of the SM particles in the s-channels or into a pair of CP-even scalars are p-wave suppressed,  there is no bound from indirect detection on these models. 

In Table~\ref{ann-real}, we show the averaged annihilation cross sections at freeze-out and the relic density for dark matter with real-scalar mediator in some benchmark models considered in Table~\ref{BRa}, having passed the diphoton condition as well as the above current collider bounds.  
Models A and B (C) belong to the left(right) plot in Fig.~\ref{relic2}.

\begin{table}[ht]
\centering
\small
\begin{tabular}{|c||c|c|c|c|c|c|}
\hline 
Model & $\langle\sigma v_{\rm rel}\rangle_{s,\gamma\gamma}$ & $\langle\sigma v_{\rm rel}\rangle_{s,gg}$ & $\langle\sigma v_{\rm rel}\rangle_{s,Z\gamma}$ & $\langle\sigma v_{\rm rel}\rangle_{s,ZZ}$ & $\langle\sigma v_{\rm rel}\rangle_{ss}$ & $\Omega_\chi h^2$   \\ [0.7ex] 
\hline 
A & $2.79\times 10^{-29}$ & $4.95\times 10^{-26}$ & $1.61\times 10^{-29}$ & $2.28\times 10^{-30}$ & $-$ & $0.116$ \\ [0.7ex]
\hline
B & $2.12\times 10^{-30}$ & $3.99\times 10^{-27}$ & $1.27\times 10^{-30}$ & $1.87\times 10^{-31}$ & $4.15\times 10^{-26}$ & $0.121$ \\ [0.7ex]
\hline
C & $2.97\times 10^{-26}$ & $6.44\times 10^{-29}$ & $1.61\times 10^{-26}$ & $2.08\times 10^{-27}$ & $-$ & $0.120$ \\ [0.7ex]
\hline
\end{tabular}
\caption{Averaged annihilation cross sections (in units of ${\rm cm^3/s}$) at freeze-out and relic density for dark matter with real-scalar. The benchmark models are the same as in Table~\ref{BRs} and Fig~\ref{relic2}. All the constraints from the current collider and cosmic data are satisfied.   }
\label{ann-real}
\end{table}

\section{Dark matter with pseudo- and real-scalars}

In this section, we consider alternative interpretations of the diphoton excess as the degenerate real- and pseudo-scalar resonances or the cascade decay of the real-scalar into a pair of pseudo-scalars, each of which decays into a pair of photons.   In these cases, we incorporate the constraints from dark matter and  collider searches in the model.

\subsection{Dark matter annihilation}

When both pseudo-scalar and real scalar are included in the effective field theory, either or both of them can produce the diphoton resonance and contribute to the annihilation of dark matter. 

\begin{figure}
  \begin{center}
   \includegraphics[height=0.15\textwidth]{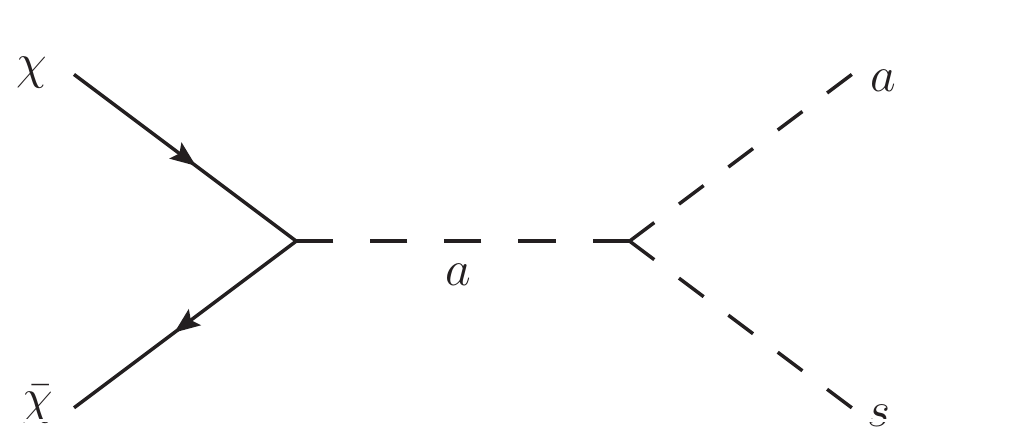}
    \includegraphics[height=0.15\textwidth]{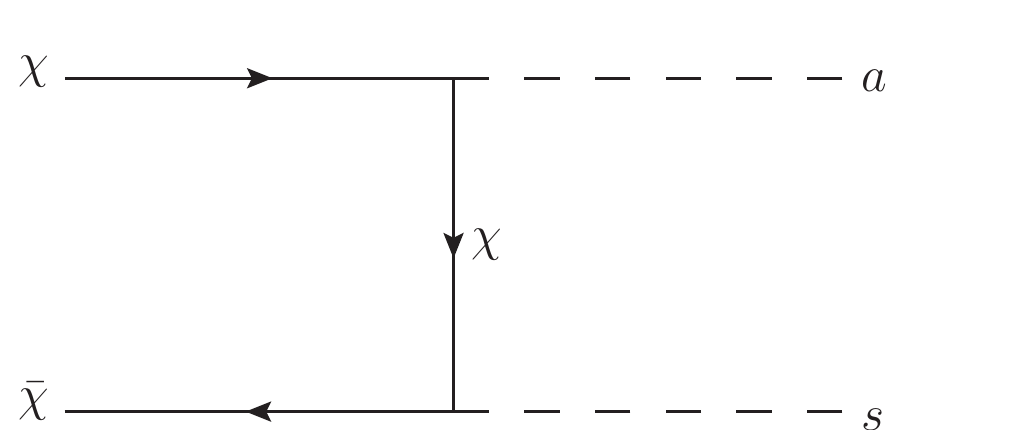}
        \includegraphics[height=0.15\textwidth]{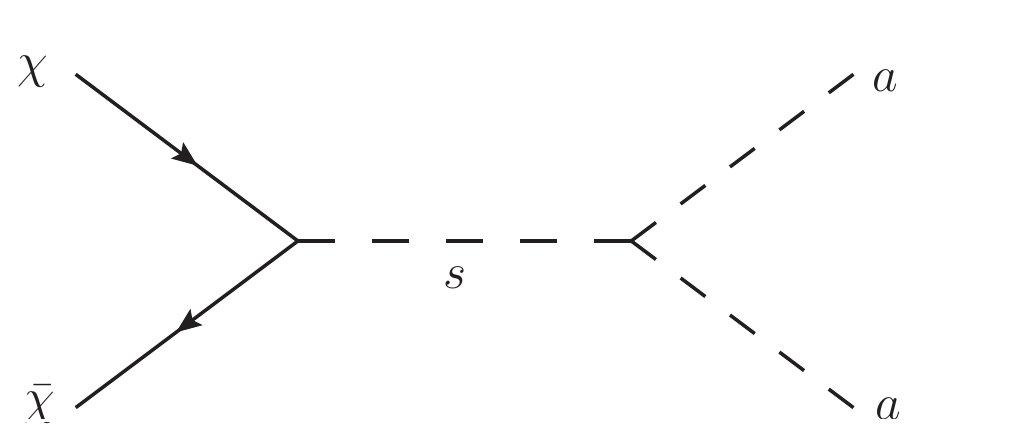}
   \end{center}
  \caption{Feynman diagrams for dark matter annihilation into a pair of real- and pseudo-scalars and an additional channel with real-scalar mediator.}
  \label{as-diagrams}
\end{figure}

For $m_\chi > (m_a + m_s)/2$, the dark matter annihilation into $as$ is open as shown in Fig.~\ref{as-diagrams} and it gives rise to an additional annihilation cross section of dark matter, given by   
\bea
(\sigma v_{\rm rel})_{as}&=& \frac{\lambda^2_\chi}{64\pi m^2_\chi}\bigg[\frac{\lambda^2_\chi(4m^2_\chi-m^2_s+m^2_a)^2}{(m^2_s+m^2_a-4m^2_\chi)^2}+\frac{8\lambda^2_S v^2_s m^2_\chi}{(4m^2_\chi-m^2_a)^2+\Gamma^2_a m^2_a} \nonumber \\
&&+\frac{4\sqrt{2}\lambda_\chi\lambda_S v_s m_\chi(4m^2_\chi-m^2_a)}{(4m^2_\chi-m^2_a)^2+\Gamma^2_a m^2_a}\cdot\frac{4m^2_\chi-m^2_s+m^2_a}{m^2_s+m^2_a-4m^2_\chi} \bigg] \nonumber \\
&&\quad \times \Big(1-\frac{(m_s-m_a)^2}{4m^2_\chi}\Big)^{1/2} \Big(1-\frac{(m_s+m_a)^2}{4m^2_\chi}\Big)^{1/2}\,.
\eea
We can see that the $as$ channel is s-wave so it is also relevant for indirect detection at present. 
In this case, the total annihilation cross section of dark matter is given by $(\sigma v_{\rm rel})_{\rm tot}=(\sigma v_{\rm rel})_a+(\sigma v_{\rm rel})_s+(\sigma v_{\rm rel})_{as}$, where the first two contributions are given in the previous sections with a single scalar resonance.

First, when two singlet scalars are almost degenerate in mass, namely, $m_a\approx m_s\approx 750\,{\rm GeV}$, they both contribute to the diphoton excesses. In this case, the new $as$ annihilation channel of dark matter is open only for a heavy dark matter with $m_\chi\gtrsim 750\,{\rm GeV}$.

On the other hand, when the pseudo-scalar or real-scalar is light enough, $m_a\lesssim 0.4\,{\rm GeV}\ll m_s=750\,{\rm GeV}$ or $m_s\lesssim 0.4\,{\rm GeV}\ll m_a=750\,{\rm GeV}$,  we can identify the real scalar or pseudo-scalar as the diphoton resonance and obtain the diphoton excess from the cascade decay of the real scalar ($s\rightarrow aa$ with $a\rightarrow \gamma\gamma$) in the former case or the direct decay of the pseudo-scalar ($a\rightarrow \gamma\gamma$) in the latter case. In these cases, the $as$ annihilation channel of dark matter is open even for a relatively light dark matter with $m_\chi\gtrsim 375\,{\rm GeV}$.  
For our discussion, we focus on the former case when the pseudo-scalar is much lighter than the real-scalar, as it is natural for a small soft breaking of the $U(1)$ global symmetry.

\subsection{Indirect detection}

\begin{figure}
  \begin{center}
   \includegraphics[height=0.40\textwidth]{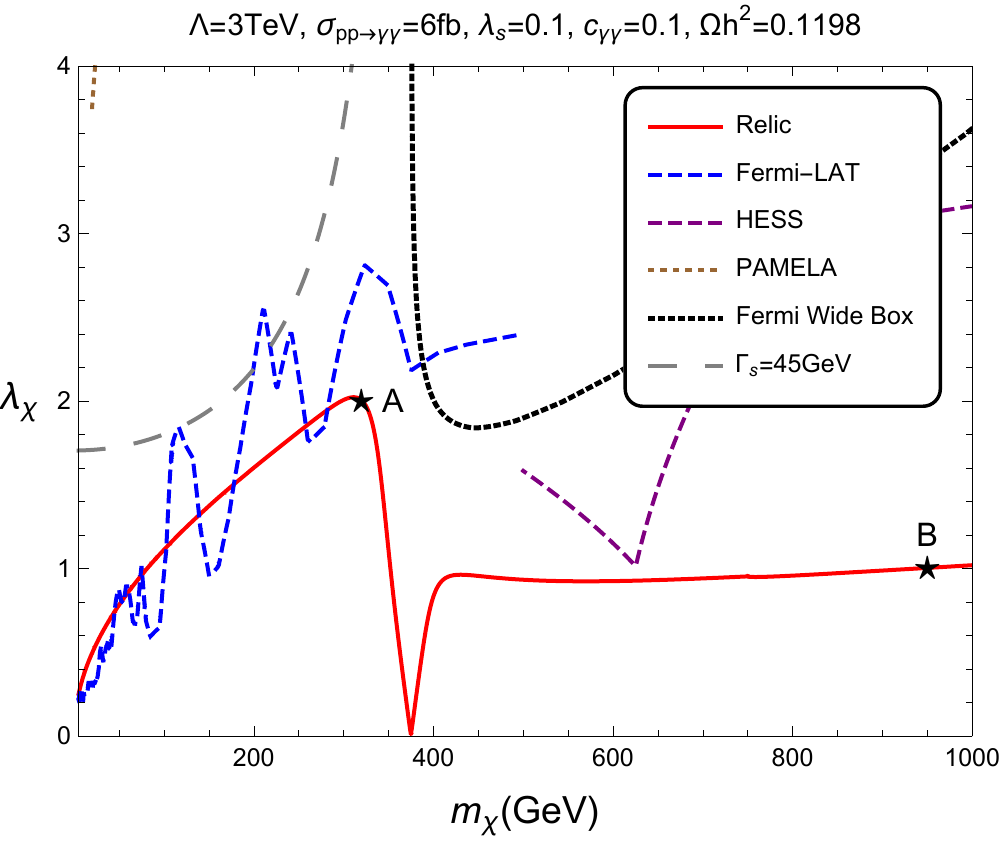}
    \includegraphics[height=0.40\textwidth]{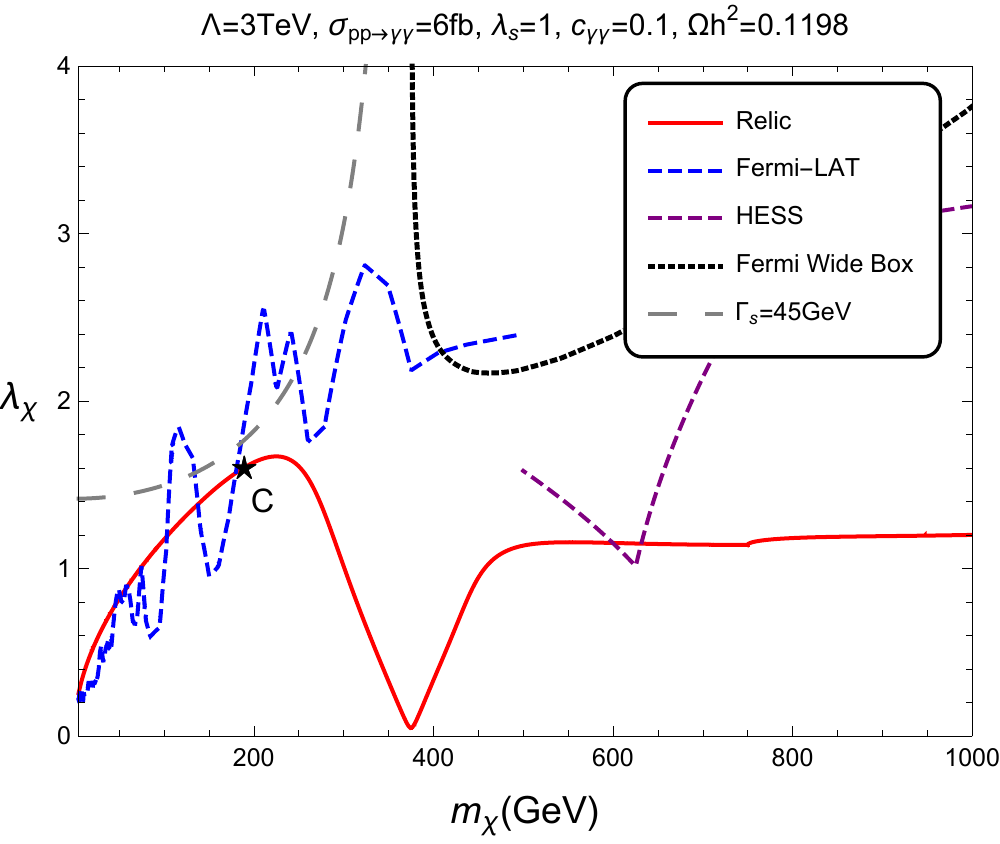} \\
      \includegraphics[height=0.40\textwidth]{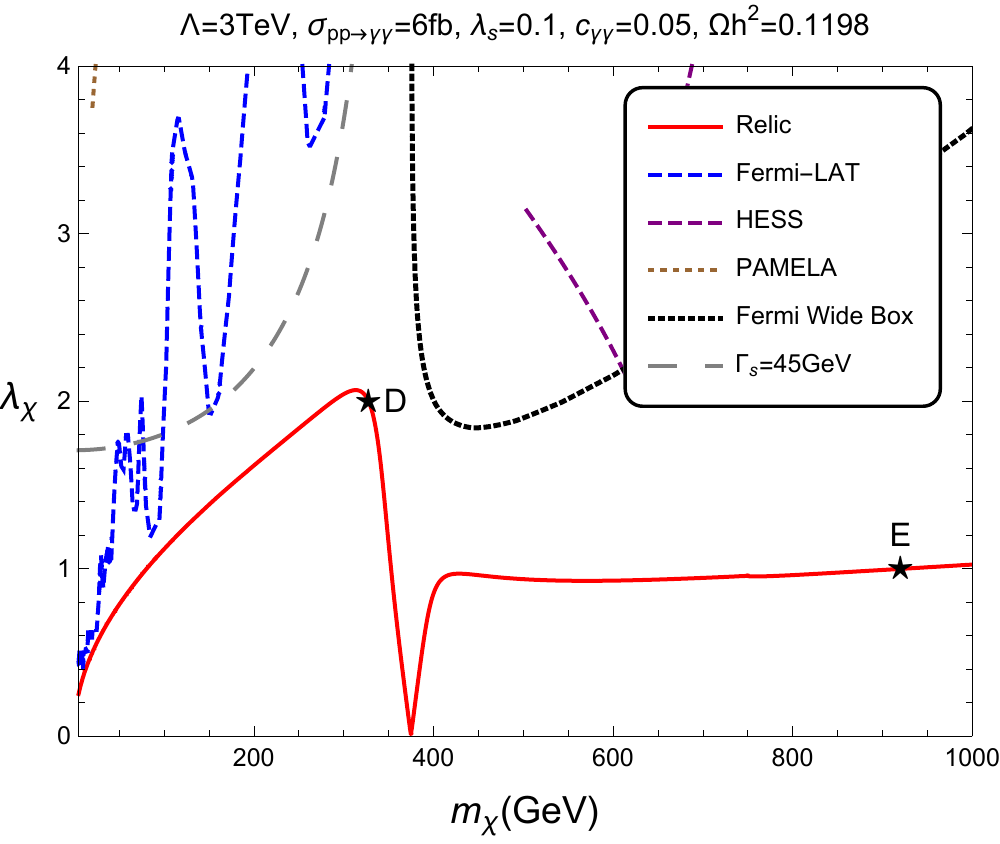}
    \includegraphics[height=0.40\textwidth]{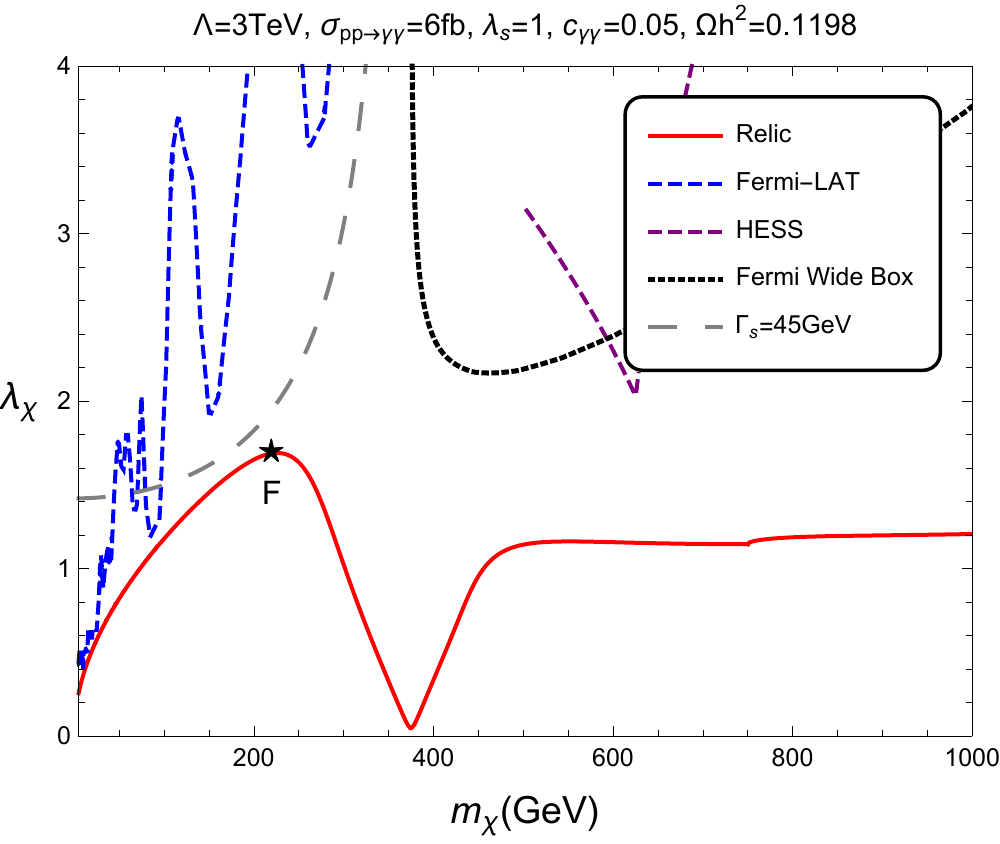} 
   \end{center}
  \caption{Parameter space of $m_\chi$ and $\lambda_\chi$ with both direct and cascade decays of the real scalar, satisfying the relic density in red lines. The condition explaining the diphoton resonance at $750\,{\rm GeV}$ for $\sigma(pp\rightarrow\gamma\gamma)=6\,{\rm fb}$ is imposed. We took $m_s=750\,{\rm GeV}$ and $m_a=0.4\,{\rm GeV}$. $c_{\gamma\gamma}=0.1$ and $\lambda_S=0.1(1.0)$ are chosen on left (right) in the upper panel while $c_{\gamma\gamma}=0.05$ and $\lambda_S=0.1(1.0)$ are chosen on left (right) in the lower panel. We have taken $c_2=0$ in all plots.
   The mono-jet limit from LHC $8\,{\rm TeV}$ and the limits from Fermi-LAT, PAMELA and HESS are shown in orange dot-dashed, blue dashed, brown dotted, and purple dashed lines, respectively. The limits from AMS-02 and Fermi-LAT wide box are also shown in pink dotted and black dotted lines. The line for $\Gamma_s=45\,{\rm GeV}$ is also shown in dashed gray. Benchmark models A and B (C) are shown in star on the left(right) plot in the upper panel, while models D and E (F) are shown in star on the left(right) plot in the lower panel. Those benchmark models are taken in Table~\ref{BRs-two}. }
  \label{relic3}
\end{figure}

As the pseudo-scalar is light, it mediates dark matter annihilations. In particular, dark matter annihilation channels,  $\chi{\bar\chi}\rightarrow a\rightarrow \gamma\gamma, Z\gamma$, are s-wave and they lead to monochromatic photons at $E_\gamma=m_\chi$ and $E_\gamma=m_\chi \Big(1-\frac{4m^2_Z}{m^2_\chi} \Big)$, respectively, as in the case with the pseudo-scalar resonance, so the model can be constrained by Ferm-LAT \cite{fermilat2} and HESS \cite{hess2013} line searches. 
 
Furthermore, annihilation channels of dark matter into $WW, ZZ, gg$ in s-channels with pseudo-scalar or the annihilation channel, $\chi{\bar\chi}\rightarrow as$ with $s\rightarrow WW, ZZ, gg$ and/or $a\rightarrow gg$ lead to continuum photons from bremstrahlung or decay and they are constrained by Fermi-diffuse gamma-ray searches from dwarf galaxies \cite{dwarfgalaxy}. 
Moreover, dark matter annihilation into gluons can be constrained by anti-proton data from PAMELA and AMS-02 \cite{antiproton}.  Since the pseudo-scalar has sub-GeV mass, the $s$-wave annihilation of weak-scale dark matter is not enhanced due to a resonance, but rather it gets smaller as dark matter increases. Furthermore, small effective couplings of scalars are allowed in the case of cascade decay. 
Therefore, the indirect bounds on the $s$-wave channels are weaker than the case with pseudo-scalar resonance.
In this case, the $p$-wave annihilation of dark matter with real-scalar resonance becomes important at freeze-out, determining the relic density. 

In the presence of a light pseudo-scalar, there is an additional $s$-wave annihilation channel,  $\chi{\bar\chi}\rightarrow as$, is $s$-wave, and it leads to multi-photons due to the direct decay $a\rightarrow \gamma\gamma$ or the cascade decay of the real scalar, $s\rightarrow aa$, with $a\rightarrow \gamma\gamma$. The gamma-ray boxes  could be constrained further by line-like features in Fermi-LAT and HESS, leading to more stringent bounds than Fermi-LAT diffuse gamma-ray searches or anti-proton searches, depending on the branching fractions of scalars.

We briefly discuss the gamma-ray energy obtained from $\chi{\bar\chi}\rightarrow a s$ channel. 
The decay $a\rightarrow \gamma \gamma$ produces two photons with identical energy in the rest frame of the pseudo-scalar, $E^*_\gamma=m_a/2$. However, in the galactic frame, where the dark matter particles move non-relativistically, the photon energy reads \cite{gbox,ibarra}
\bea
E_\gamma=\frac{1}{\gamma_a} E^*_{\gamma} (1-v_a \cos\theta_a )^{-1}\,,
\eea
where $\gamma_a\equiv 1/\sqrt{1-v^2_a}$, $\theta_a$ is the angle between the direction of the pseudo-scalar and the direction of the photon and $v_a$ is the pseudo-scalar velocity, given by
\be
v_a=\frac{p_a}{E_a}=\sqrt{1-\frac{m^2_a}{m^2_\chi}\Big(1 +\frac{m^2_a-m^2_s}{4m_\chi^2}\Big)^{-2}}\,.
\ee
Since the pseudo-scalar decays isotropically, the resulting energy spectrum presents a box-shaped structure with the photon energy ranging from $E_-$ to $E_+$, where $E_\pm=\frac{1}{2}A m_\chi(1\pm \sqrt{1-\frac{m^2_a}{A^2m^2_\chi} })$ and $A=1+(m^2_a-m^2_s)/(4 m^2_\chi)$.
Then, the energy spectrum of two hard photons produced in the annihilation channel $\chi {\bar \chi}\rightarrow a s$ \cite{ibarra} is
\begin{eqnarray}
\frac{dN^{(2)}_\gamma}{dE_\gamma}&=&\frac{2}{E_+-E_-}\Theta(E_\gamma-E_-)\Theta(E_+-E_\gamma) {\rm BR}(a\rightarrow \gamma\gamma)
\label{eq:spectrum-as}
\end{eqnarray}
where $\Theta$ is the Heaviside function. 

Furthermore, the cascade decay $s\rightarrow aa\rightarrow 4\gamma$ with a large ${\rm BR}(a\rightarrow\gamma\gamma)$ leads to four additional photons, thus leading to potentially interesting signatures in gamma-ray searches, which will be published elsewhere. 
In this work, we focus on the box-shaped gamma-ray spectrum to get a conservative bound on the annihilation cross section for $\chi{\bar\chi}\rightarrow as$.

\begin{table}[ht]
\centering
\small
\begin{tabular}{|c||c|c|c|c|}
\hline 
Model & $\langle\sigma v_{\rm rel}\rangle_{\gamma\gamma}$ & $\langle\sigma v_{\rm rel}\rangle_{gg}$ & $\langle\sigma v_{\rm rel}\rangle_{Z\gamma}$ & $\langle\sigma v_{\rm rel}\rangle_{ZZ}$\\ [0.7ex] 
\hline 
A & $2.06\times 10^{-27}$ & $1.25\times 10^{-28}$ & $1.17\times 10^{-27}$ & $1.64\times 10^{-28}$ \\ [0.7ex]
\hline
B & $5.16\times 10^{-28}$ & $4.98\times 10^{-30}$ & $3.10\times 10^{-28}$ & $4.58\times 10^{-29}$ \\ [0.7ex]
\hline
C & $1.32\times 10^{-27}$ & $3.27\times 10^{-29}$ & $6.69\times 10^{-28}$ & $8.03\times 10^{-29}$ \\ [0.7ex]
\hline
D & $5.16\times 10^{-28}$ & $1.07\times 10^{-28}$ & $2.95\times 10^{-28}$ & $4.12\times 10^{-29}$ \\ [0.7ex]
\hline
E & $1.29\times 10^{-28}$ & $4.88\times 10^{-30}$ & $7.75\times 10^{-29}$ & $1.14\times 10^{-29}$ \\ [0.7ex]
\hline
F & $3.73\times 10^{-28}$ & $3.55\times 10^{-29}$ & $1.98\times 10^{-28}$ & $2.53\times 10^{-29}$ \\ [0.7ex]
\hline
\end{tabular}
\end{table}  

\begin{table}[ht]
\centering
\small
\begin{tabular}{|c||c|c|c|c|}
\hline 
Model & $\langle\sigma v_{\rm rel}\rangle_{aa}$ & $\langle\sigma v_{\rm rel}\rangle_{as}$ & $\langle\sigma v_{\rm rel}\rangle_{ss}$ & $\Omega_\chi h^2$\\ [0.7ex] 
\hline 
A & $8.47\times 10^{-26}$ & $-$ & $-$ & $0.121$ \\ [0.7ex]
\hline
B & $1.10\times 10^{-27}$ & $4.16\times 10^{-26}$ & $3.19\times 10^{-27}$ & $0.122$ \\ [0.7ex]
\hline
C & $9.14\times 10^{-26}$ & $-$ & $-$ & $0.121$ \\ [0.7ex]
\hline
D & $9.31\times 10^{-26}$ & $-$ & $-$ & $0.119$ \\ [0.7ex]
\hline
E & $1.18\times 10^{-27}$ & $4.34\times 10^{-26}$ & $3.45\times 10^{-27}$ & $0.119$ \\ [0.7ex]
\hline
F & $9.76\times 10^{-26}$ & $-$ & $-$ & $0.117$ \\ [0.7ex]
\hline
\end{tabular}
\caption{Averaged annihilation cross sections (in units of ${\rm cm^3/s}$) at present and relic density for dark matter with two scalars, except that those for $aa, ss$ channels are given at freeze-out.
The benchmark models are the same as in Table~\ref{BRs-two} and Fig.~\ref{relic3}. All the constraints from the current collider and cosmic data are satisfied. }
\label{ann-as}
\end{table}

In Fig.~\ref{relic3}, we show the parameter space of dark matter mass $m_\chi$ and coupling $\lambda_\chi$ in the model with $m_s=750\,{\rm GeV}$ and $m_a=0.4\,{\rm GeV}$ where the condition for diphoton excesses is satisfied. We have set $c_2=0$ for simplicity. 
Depending on the value of the gluon coupling $c_{gg}=0.1 (0.01)$ in the upper and lower panels, respectively, with the photon coupling $c_{\gamma\gamma}$ being determined by the diphoton condition (\ref{diphoton-a}), or the quartic coupling of the complex scalar, $\lambda_S=0.1 (1.0)$ on left and right in each panel, respectively, we imposed the current bounds from mono-jet searches as well as various indirect detections.  

First, on the left plot in the upper panel with $c_{\gamma\gamma}=\lambda_S=0.1$ in Fig.~\ref{relic3}, the mono-jet bound excludes most of the region below resonance, while Fermi-LAT line and other indirect searches are not sensitive enough yet to constrain the region with saturated relic density. When the photon coupling gets smaller but the quartic coupling remains small as on the left plot in the lower panel with $c_{\gamma\gamma}=0.05$ and $\lambda_S=0.1$, the Fermi-LAT line search does not constrain the region below resonance. 
The important difference from the case with pseudo-scalar resonance is that there appears a bound from the Fermi-LAT search for gamma-ray box, although it is not sensitive enough yet to the region with saturated relic density.  
Finally, when the quartic coupling gets larger to $\lambda_S=1.0$ as in the right plots with $c_{\gamma\gamma}=0.1$ or $0.05$, the diphoton excesses can be explained dominantly by the cascade decay of real-scalar. In these cases, either mono-jet or Fermi-LAT searches do not reach the region with saturated relic density, opening up more parameter space to be probed for in the LHC Run 2 and future gamma-ray searches such as Cherenkov Telescope Array \cite{CTA}.

In Table~\ref{ann-as}, we show the averaged annihilation cross sections for $s$-wave channels such as $\gamma\gamma, gg, Z\gamma, ZZ, as$, at present, except those for $aa, ss$ channels, which are taken at freeze-out, and the relic density for dark matter with two scalar fields in some benchmark models considered in Table~\ref{BRs-two}. The diphoton condition as well as the above current collider bounds are satisfied for all the models in Table~\ref{ann-as}.  We have set the scalar masses to $m_s=750\,{\rm GeV}$ and $m_a=0.4\,{\rm GeV}$ and the effective gauge couplings to $d_i=\frac{4}{3}c_i (i=1,3)$ and $c_2=d_2=0$. Models A and B (C) belong to the left(right) plot in the upper panel of Fig.~\ref{relic3}, while Model E and D (F) belongs to the left(right) plot in the lower panel of Fig.~\ref{relic3}.
These models satisfy the current bounds from various indirect detection experiments discussed above. 
As the $aa, ss$ channels are $p$-wave suppressed and negligible at present, there is no bound on those channels from indirect detection. So, we show in the same table the annihilation cross sections for the 
$aa, ss$ channels at freeze-out.

\section{Conclusions}

We have considered various possibilities of explaining the diphoton excesses observed at the LHC in terms of singlet scalar resonances with effective interactions to gluons and photon.  In the case that the resonance decays directly into a photon pair, the region of the parameter space where there is an invisible decay of the resonance into a pair of dark matter particles is strongly constrained by the interplay between mono-jet and Fermi-LAT gamma-ray searches. When the diphoton excesses stem from the cascade decay of the real-scalar into a pair of pseudo-scalars, the effective couplings for SM gauge bosons  can be smaller. In this case, the collider and indirect detection bounds are less strong, but the gamma-ray box coming from the cascade annihilation of dark matter into a pair of real-scalar and pseudo-scalar could be a smoking-gun signal in gamma-ray searches.
We have shown various benchmark models that are consistent with all the collider and astrophysical constraints and can be testable in the LHC Run 2 as well as future gamma-ray searches.



\section*{Acknowledgments}

The work of HML is supported in part by Basic Science Research Program through the National Research Foundation of Korea (NRF) funded by the Ministry of Education, Science and Technology (2013R1A1A2007919). The work of YJK is supported by the Chung-Ang University Graduate Research Scholarship in 2016.

\def\theequation{A.\arabic{equation}}

\setcounter{equation}{0}

\vskip0.8cm
\noindent
{\Large \bf Appendix A:  Scalar sector of the model}
\vskip0.4cm
\noindent

In the text, we consider the scalar potential for the singlet complex scalar $S$ and the SM Higgs doublet $H$ is
\bea
V(H,S)= \lambda_H |H|^4+\lambda_S |S|^4+2\lambda_{HS}|S|^2 |H|^2 +m^2_H |H|^2+m^2_S|S|^2-\Big(\frac{1}{2}m^{\prime 2}_S S^2+{\rm h.c.}\Big) \label{potential}
\eea
where $m'_S$ term breaks the $U(1) $ global symmetry softly to give the pseudo-scalar component mass.
After minimizing the potential in eq.(\ref{potential}), the VEVs of the singlet and the Higgs doublet, $v_s$ and $v$, are determined as
\bea
v^2_s&=&  \frac{\lambda_{HS} m^2_H-\lambda_H (m^2_S-m^{\prime 2}_S)}{\lambda_S\lambda_H-\lambda^2_{HS}},\\
v^2&=&  \frac{\lambda_{HS}(m^2_S-m^{'2}_S)-\lambda_S m^2_H}{\lambda_S\lambda_H-\lambda^2_{HS}}.
\eea
The conditions for a local minimum are $\lambda_{HS}m^2_H-\lambda_H (m^2_S-m^{\prime 2}_S)>0$, $\lambda_{HS}(m^2_S-m^{\prime 2}_S)-\lambda_S m^2_H>0$ and $\lambda_S \lambda_H-\lambda^2_{HS}>0$.
Expanding the scalar fields around the vacuum as $S=(v_s+s+ia)/\sqrt{2}$ and  $H^T=(0,v+h)/\sqrt{2}$ in unitary gauge,
the obtained mass matrix for CP-even scalars can be diagonalized by the field rotation,
\be
s=\cos\theta\, h_2+\sin\theta\,h_1, \quad h=-\sin\theta\, h_2+\cos\theta\,h_1
\ee
with
\be
\tan\,2\theta=\frac{2\lambda_{HS} v_s v}{\lambda_H v^2-\lambda_S v^2_s},
\ee
and the mass eigenvalues are
\bea
m^2_{1,2}=\lambda_H v^2+ \lambda_S v^2_s\mp \sqrt{(\lambda_S v^2_s-\lambda_H v^2)^2+4\lambda^2_{HS} v^2 v^2_s}. \label{masses}
\eea
Thus, $h_1$ is Higgs-like and $h_2$ is singlet-like.

We also note that the singlet-like scalar $h_2$ can have Higgs-like couplings to the SM particles through the mixing with the Higgs boson as well as scalar triple self-couplings given by 
\bea
{\cal L}_{\rm scalar}&=& c_{h_1 aa} h_1 a^2 + c_{ h_2aa} h^2_2 a^2   \nonumber \\
&&+ c_{h_1 h_1 h_1} h^3_1+c_{h_2 h_2 h_2} h^3_2 + c_{h_1 h_2 h_2} h_1 h^2_2 +c_{h_1 h_1 h_2} h^2_1 h_2  \label{scalar}
\eea
 where
 \bea
 c_{h_1 aa}&=&  -\lambda_{HS}v\cos\theta -\lambda_S v_s \sin\theta, \\
 c_{h_2 aa}&=&  \lambda_{HS} v\sin\theta - \lambda_S v_s \cos\theta, \\
 c_{h_1 h_1 h_1}  &=& -\lambda_H v\cos^3\theta-\lambda_{HS}\cos\theta\sin\theta (v\sin\theta+v_s\cos\theta)-\lambda_S v_s \sin^3\theta,\\
 c_{h_2 h_2 h_2}  &=& \lambda_H v\sin^3\theta+\lambda_{HS}\cos\theta\sin\theta (v\cos\theta-v_s \sin\theta)-\lambda_S v_s\cos^3\theta, \\
 c_{h_1 h_2 h_2} &=&-3\lambda_H v\sin^2\theta\cos\theta-\lambda_{HS}\Big(v\cos\theta(\cos^2\theta-2\sin^2\theta ) \nonumber \\
 &&+v_s\sin\theta (\sin^2\theta -2\cos^2\theta) \Big) 
-3\lambda_S v_s\sin\theta \cos^2\theta, \\
  c_{h_1 h_1 h_2} &=&3\lambda_H v\cos^2\theta\sin\theta+\lambda_{HS}\Big(v\sin\theta(\sin^2\theta-2\cos^2\theta ) \nonumber \\
  &&-v_s\cos\theta (\cos^2\theta -2\sin^2\theta) \Big)
 -3\lambda_S v_s\cos\theta \sin^2\theta.
 \eea


\begin{thebibliography}{999}


\bibitem{diphoton}
ATLAS collaboration,
  ATLAS-CONF-2015-081;
  CMS Collaboration,
  CMS-PAS-EXO-15-004.


\bibitem{moriond}
  ATLAS collaboration,
  ATLAS-CONF-2016-018;
  CMS Collaboration,
  CMS-PAS-EXO-16-018.


\bibitem{scalar}
  R.~Franceschini {\it et al.},
  arXiv:1512.04933 [hep-ph];
  R.~Franceschini, G.~F.~Giudice, J.~F.~Kamenik, M.~McCullough, F.~Riva, A.~Strumia and R.~Torre,
  arXiv:1604.06446 [hep-ph].

\bibitem{production}
  J.~Ellis, S.~A.~R.~Ellis, J.~Quevillon, V.~Sanz and T.~You,
  JHEP {\bf 1603} (2016) 176
  doi:10.1007/JHEP03(2016)176
  [arXiv:1512.05327 [hep-ph]];
  A.~Falkowski, O.~Slone and T.~Volansky,
  JHEP {\bf 1602} (2016) 152
  doi:10.1007/JHEP02(2016)152
  [arXiv:1512.05777 [hep-ph]];
  J.~S.~Kim, K.~Rolbiecki and R.~Ruiz de Austri,
  Eur.\ Phys.\ J.\ C {\bf 76} (2016) no.5,  251
  doi:10.1140/epjc/s10052-016-4102-0
  [arXiv:1512.06797 [hep-ph]];
  M.~R.~Buckley,
  arXiv:1601.04751 [hep-ph];
  J.~F.~Kamenik, B.~R.~Safdi, Y.~Soreq and J.~Zupan,
  arXiv:1603.06566 [hep-ph].



\bibitem{scalar2}
  K.~Harigaya and Y.~Nomura,
  Phys.\ Lett.\ B {\bf 754} (2016) 151
  doi:10.1016/j.physletb.2016.01.026
  [arXiv:1512.04850 [hep-ph]];
  A.~Pilaftsis,
  Phys.\ Rev.\ D {\bf 93} (2016) no.1,  015017
  doi:10.1103/PhysRevD.93.015017
  [arXiv:1512.04931 [hep-ph]];
  T.~Higaki, K.~S.~Jeong, N.~Kitajima and F.~Takahashi,
  Phys.\ Lett.\ B {\bf 755} (2016) 13
  doi:10.1016/j.physletb.2016.01.055
  [arXiv:1512.05295 [hep-ph]];
  arXiv:1512.05700 [hep-ph];
  A.~Belyaev, G.~Cacciapaglia, H.~Cai, T.~Flacke, A.~Parolini and H.~Serôdio,
  arXiv:1512.07242 [hep-ph];
  L.~J.~Hall, K.~Harigaya and Y.~Nomura,
  JHEP {\bf 1603} (2016) 017
  doi:10.1007/JHEP03(2016)017
  [arXiv:1512.07904 [hep-ph]];
  M.~Son and A.~Urbano,
  arXiv:1512.08307 [hep-ph];
  J.~E.~Kim,
  Phys.\ Lett.\ B {\bf 755} (2016) 190
  doi:10.1016/j.physletb.2016.02.016
  [arXiv:1512.08467 [hep-ph]];
  A.~Angelescu, A.~Djouadi and G.~Moreau,
  Phys.\ Lett.\ B {\bf 756} (2016) 126
  doi:10.1016/j.physletb.2016.02.064
  [arXiv:1512.04921 [hep-ph]];
  S.~Di Chiara, L.~Marzola and M.~Raidal,
  arXiv:1512.04939 [hep-ph];
  W.~Altmannshofer, J.~Galloway, S.~Gori, A.~L.~Kagan, A.~Martin and J.~Zupan,
  arXiv:1512.07616 [hep-ph];
  P.~Ko, Y.~Omura and C.~Yu,
  JHEP {\bf 1604} (2016) 098
  doi:10.1007/JHEP04(2016)098
  [arXiv:1601.00586 [hep-ph]];
  T.~Modak, S.~Sadhukhan and R.~Srivastava,
  Phys.\ Lett.\ B {\bf 756} (2016) 405
  doi:10.1016/j.physletb.2016.03.021
  [arXiv:1601.00836 [hep-ph]];
  C.~W.~Chiang, H.~Fukuda, M.~Ibe and T.~T.~Yanagida,
  arXiv:1602.07909 [hep-ph];
  C.~Bonilla, M.~Nebot, R.~Srivastava and J.~W.~F.~Valle,
  Phys.\ Rev.\ D {\bf 93} (2016) no.7,  073009
  doi:10.1103/PhysRevD.93.073009
  [arXiv:1602.08092 [hep-ph]];
  H.~P.~Nilles and M.~W.~Winkler,
  arXiv:1604.03598 [hep-ph];
  K.~Choi, S.~H.~Im, H.~Kim and D.~Y.~Mo,
  arXiv:1605.00206 [hep-ph];
  A.~Djouadi and A.~Pilaftsis,
  arXiv:1605.01040 [hep-ph].


\bibitem{gmdm-diphoton}
  C.~Han, H.~M.~Lee, M.~Park and V.~Sanz,
  Phys.\ Lett.\ B {\bf 755} (2016) 371
  doi:10.1016/j.physletb.2016.02.040
  [arXiv:1512.06376 [hep-ph]].
  
 \bibitem{spin2}
  S.~B.~Giddings and H.~Zhang,
  arXiv:1602.02793 [hep-ph];
  A.~Falkowski and J.~F.~Kamenik,
  arXiv:1603.06980 [hep-ph];
  J.~L.~Hewett and T.~G.~Rizzo,
  arXiv:1603.08250 [hep-ph];
  B.~M.~Dillon and V.~Sanz,
  arXiv:1603.09550 [hep-ph].


\bibitem{coexist}
  V.~Sanz,
  arXiv:1603.05574 [hep-ph];
  P.~Roig and J.~J.~Sanz-Cillero,
  arXiv:1605.03831 [hep-ph].


\bibitem{diphotondm}
  Y.~Mambrini, G.~Arcadi and A.~Djouadi,
  Phys.\ Lett.\ B {\bf 755} (2016) 426
  doi:10.1016/j.physletb.2016.02.049
  [arXiv:1512.04913 [hep-ph]];
  M.~Backovic, A.~Mariotti and D.~Redigolo,
  JHEP {\bf 1603} (2016) 157
  doi:10.1007/JHEP03(2016)157
  [arXiv:1512.04917 [hep-ph]];
  X.~J.~Bi, Q.~F.~Xiang, P.~F.~Yin and Z.~H.~Yu,
  Nucl.\ Phys.\ B {\bf 909} (2016) 43
  doi:10.1016/j.nuclphysb.2016.04.042
  [arXiv:1512.06787 [hep-ph]];
  U.~K.~Dey, S.~Mohanty and G.~Tomar,
  Phys.\ Lett.\ B {\bf 756} (2016) 384
  doi:10.1016/j.physletb.2016.03.048
  [arXiv:1512.07212 [hep-ph]];
  J.~C.~Park and S.~C.~Park,
  arXiv:1512.08117 [hep-ph].
  K.~Ghorbani and H.~Ghorbani,
  arXiv:1601.00602 [hep-ph];
  S.~Bhattacharya, S.~Patra, N.~Sahoo and N.~Sahu,
  arXiv:1601.01569 [hep-ph];
  F.~D'Eramo, J.~de Vries and P.~Panci,
  arXiv:1601.01571 [hep-ph];
  P.~Ko and T.~Nomura,
  arXiv:1601.02490 [hep-ph];
  A.~Salvio, F.~Staub, A.~Strumia and A.~Urbano,
  JHEP {\bf 1603} (2016) 214
  doi:10.1007/JHEP03(2016)214
  [arXiv:1602.01460 [hep-ph]];
  S.~F.~Ge, H.~J.~He, J.~Ren and Z.~Z.~Xianyu,
  Phys.\ Lett.\ B {\bf 757} (2016) 480
  doi:10.1016/j.physletb.2016.04.008
  [arXiv:1602.01801 [hep-ph]];
  M.~Redi, A.~Strumia, A.~Tesi and E.~Vigiani,
  arXiv:1602.07297 [hep-ph];
  E.~Morgante, D.~Racco, M.~Rameez and A.~Riotto,
  arXiv:1603.05592 [hep-ph];
  G.~Arcadi, P.~Ghosh, Y.~Mambrini and M.~Pierre,
  arXiv:1603.05601 [hep-ph].


\bibitem{cascade}
  S.~Knapen, T.~Melia, M.~Papucci and K.~Zurek,
  Phys.\ Rev.\ D {\bf 93} (2016) no.7,  075020
  doi:10.1103/PhysRevD.93.075020
  [arXiv:1512.04928 [hep-ph]];
  P.~Agrawal, J.~Fan, B.~Heidenreich, M.~Reece and M.~Strassler,
  arXiv:1512.05775 [hep-ph];
  J.~Chang, K.~Cheung and C.~T.~Lu,
  arXiv:1512.06671 [hep-ph];
  L.~Aparicio, A.~Azatov, E.~Hardy and A.~Romanino,
  arXiv:1602.00949 [hep-ph];
  U.~Ellwanger and C.~Hugonie,
  arXiv:1602.03344 [hep-ph];
  F.~Domingo, S.~Heinemeyer, J.~S.~Kim and K.~Rolbiecki,
  arXiv:1602.07691 [hep-ph];
  V.~De Romeri, J.~S.~Kim, V.~Martin-Lozano, K.~Rolbiecki and R.~R.~de Austri,
  arXiv:1603.04479 [hep-ph];
  Y.~Jiang, L.~Li and R.~Zheng,
  arXiv:1605.01898 [hep-ph].



\bibitem{lpp}
  H.~M.~Lee, M.~Park and W.~I.~Park,
  Phys.\ Rev.\ D {\bf 86} (2012) 103502
  [arXiv:1205.4675 [hep-ph]];
  H.~M.~Lee, M.~Park and W.~I.~Park,
  JHEP {\bf 1212} (2012) 037
  [arXiv:1209.1955 [hep-ph]].

\bibitem{MET}
  H.~M.~Lee, M.~Park and V.~Sanz,
  JHEP {\bf 1303} (2013) 052
  [arXiv:1212.5647 [hep-ph]].




\bibitem{ibarra}
  A.~Ibarra, H.~M.~Lee, S.~Lopez Gehler, W.~I.~Park and M.~Pato,
  JCAP {\bf 1305} (2013) 016
  [arXiv:1303.6632 [hep-ph]].


\bibitem{hambye}
  X.~Chu, T.~Hambye, T.~Scarna and M.~H.~G.~Tytgat,
  Phys.\ Rev.\ D {\bf 86} (2012) 083521
  doi:10.1103/PhysRevD.86.083521
  [arXiv:1206.2279 [hep-ph]].



\bibitem{global}
  Y.~Mambrini, S.~Profumo and F.~S.~Queiroz,
  arXiv:1508.06635 [hep-ph].


\bibitem{discrete}
  H.~M.~Lee, S.~Raby, M.~Ratz, G.~G.~Ross, R.~Schieren, K.~Schmidt-Hoberg and P.~K.~S.~Vaudrevange,
  Nucl.\ Phys.\ B {\bf 850} (2011) 1
  doi:10.1016/j.nuclphysb.2011.04.009
  [arXiv:1102.3595 [hep-ph]];
  K.~Y.~Choi, E.~J.~Chun and H.~M.~Lee,
  Phys.\ Rev.\ D {\bf 82} (2010) 105028
  doi:10.1103/PhysRevD.82.105028
  [arXiv:1002.4791 [hep-ph]].


\bibitem{gmdm}
  H.~M.~Lee, M.~Park and V.~Sanz,
  JHEP {\bf 1405} (2014) 063
  doi:10.1007/JHEP05(2014)063
  [arXiv:1401.5301 [hep-ph]];
  H.~M.~Lee, M.~Park and V.~Sanz,
  Eur.\ Phys.\ J.\ C {\bf 74} (2014) 2715
  doi:10.1140/epjc/s10052-014-2715-8
  [arXiv:1306.4107 [hep-ph]].



\bibitem{atlasH}
  G.~Aad {\it et al.} [ATLAS Collaboration],
  Eur.\ Phys.\ J.\ C {\bf 76} (2016) no.1,  6
  doi:10.1140/epjc/s10052-015-3769-y
  [arXiv:1507.04548 [hep-ex]].


\bibitem{cmsH}
  V.~Khachatryan {\it et al.} [CMS Collaboration],
  Eur.\ Phys.\ J.\ C {\bf 75} (2015) no.5,  212
  doi:10.1140/epjc/s10052-015-3351-7
  [arXiv:1412.8662 [hep-ex]].
  

\bibitem{atlasg2}
  G.~Aad {\it et al.} [ATLAS Collaboration],
  Phys.\ Rev.\ D {\bf 90} (2014) no.11,  112015
  doi:10.1103/PhysRevD.90.112015
  [arXiv:1408.7084 [hep-ex]].


\bibitem{cmsg2}
  V.~Khachatryan {\it et al.} [CMS Collaboration],
  Eur.\ Phys.\ J.\ C {\bf 74} (2014) no.10,  3076
  doi:10.1140/epjc/s10052-014-3076-z
  [arXiv:1407.0558 [hep-ex]].




\bibitem{monojet}
  V.~Khachatryan {\it et al.} [CMS Collaboration],
  Eur.\ Phys.\ J.\ C {\bf 75} (2015) no.5,  235
  doi:10.1140/epjc/s10052-015-3451-4
  [arXiv:1408.3583 [hep-ex]].


\bibitem{di-jet}
  G.~Aad {\it et al.} [ATLAS Collaboration],
  Phys.\ Rev.\ D {\bf 91} (2015) no.5,  052007
  doi:10.1103/PhysRevD.91.052007
  [arXiv:1407.1376 [hep-ex]].


  
 \bibitem{jaeckel}
  J.~Jaeckel and M.~Spannowsky,
  Phys.\ Lett.\ B {\bf 753} (2016) 482
  doi:10.1016/j.physletb.2015.12.037
  [arXiv:1509.00476 [hep-ph]].
  
  
\bibitem{fermilat2}
  W.~B.~Atwood {\it et al.}  [LAT Collaboration],
  Astrophys.\ J.\  {\bf 697} (2009) 1071
  [arXiv:0902.1089 [astro-ph.IM]];
  A.~A.~Abdo {\it et al.},
  Phys.\ Rev.\ Lett.\  {\bf 104} (2010) 091302
  [arXiv:1001.4836 [astro-ph.HE]];
  F.~M.~Ackermann {\it et al.}  [LAT Collaboration],
  arXiv:1205.2739 [astro-ph.HE];
  M.~Ackermann {\it et al.} [Fermi-LAT Collaboration],
  Phys.\ Rev.\ D {\bf 91} (2015) 12,  122002
  doi:10.1103/PhysRevD.91.122002
  [arXiv:1506.00013 [astro-ph.HE]].


\bibitem{dwarfgalaxy}
  M.~Ackermann {\it et al.}  [Fermi-LAT Collaboration],
  Phys.\ Rev.\ Lett.\  {\bf 107} (2011) 241302
  [arXiv:1108.3546 [astro-ph.HE]];
  M.~Ackermann {\it et al.}  [Fermi-LAT Collaboration],
  arXiv:1310.0828 [astro-ph.HE];
  M.~Ackermann {\it et al.} [Fermi-LAT Collaboration],
  Phys.\ Rev.\ Lett.\  {\bf 115} (2015) 23,  231301
  doi:10.1103/PhysRevLett.115.231301
  [arXiv:1503.02641 [astro-ph.HE]].


\bibitem{hess2013}
  A.~Abramowski {\it et al.}  [H.E.S.S. Collaboration],
  Phys.\ Rev.\ Lett.\  {\bf 110} (2013) 041301
  [arXiv:1301.1173 [astro-ph.HE]].


\bibitem{antiproton}
  O.~Adriani {\it et al.}  [PAMELA Collaboration],
  Phys.\ Rev.\ Lett.\  {\bf 105} (2010) 121101
  [arXiv:1007.0821 [astro-ph.HE]];
  G.~Belanger, C.~Boehm, M.~Cirelli, J.~Da Silva and A.~Pukhov,
  JCAP {\bf 1211} (2012) 028
  [arXiv:1208.5009 [hep-ph]];
  AMS-02 Collaboration, Talks at AMS Days at CERN, 15-17 April, 2015;
  G.~Giesen, M.~Boudaud, Y.~G�nolini, V.~Poulin, M.~Cirelli, P.~Salati and P.~D.~Serpico,
  JCAP {\bf 1509} (2015) 09,  023
  doi:10.1088/1475-7516/2015/09/023, 10.1088/1475-7516/2015/9/023
  [arXiv:1504.04276 [astro-ph.HE]].


\bibitem{gbox}
  A.~Ibarra, S.~Lopez Gehler and M.~Pato,
  JCAP {\bf 1207} (2012) 043
  doi:10.1088/1475-7516/2012/07/043
  [arXiv:1205.0007 [hep-ph]].


\bibitem{CTA}
  A.~Ibarra, A.~S.~Lamperstorfer, S.~López-Gehler, M.~Pato and G.~Bertone,
  JCAP {\bf 1509} (2015) no.09,  048
  doi:10.1088/1475-7516/2015/09/048
  [arXiv:1503.06797 [astro-ph.HE]].


\end{thebibliography}
\end{document}